\documentclass[prologue, table, sigconf, screen, nonacm]{acmart}

\AtBeginDocument{%
  }

\setcopyright{none} 
\copyrightyear{2024}
\acmYear{2024}
\acmDOI{XXXXXXX.XXXXXXX}

\usepackage{pifont}
\usepackage{indentfirst}
\usepackage{amsmath}
\usepackage{booktabs}
\usepackage{threeparttable}
\usepackage{array}
\usepackage{multicol}
\begin{document}

\title{Efficient Architecture for RISC-V Vector Memory Access}

% \author{Hongyi Guan}
% \authornote{Hongyi Guan and Yichuan Gao have contributed equally and are considered to be co-first authors.}
% % \authornote{Work done at Intel.}
% \affiliation{
% %    \institution{Tsinghua University}
% %     % \country{China}
% }
% % \email{guankoala@outlook.com}

% \author{Yichuan Gao}
% \affiliation{
% %    \institution{Intel}
% %     % \country{China}
% }
% % \email{yichuan.gao@intel.com}

% \author{Chenlu Miao}
% \authornote{Chenlu Miao is the corresponding author.}
% % \authornotemark[2]
% \affiliation{
% %    \institution{Independent Researcher}
% %     % \country{China}
% }
% % \email{chenlu.miao@outlook.com}

% \author{Haoyang Wu}
% \affiliation{
% %    \institution{Intel}
% %     % \country{China}
% }
% % \email{haoyang.wu@intel.com}

% \author{Hang Zhu}
% \affiliation{
% %    \institution{Independent Researcher}
% %     % \country{China}
% }
% % \email{zhuhang.cn@outlook.com}
% % % \email{zhuhang18@mails.ucas.ac.cn}

% \author{Mingfeng Lin}
% \affiliation{
% %    \institution{Shenzhen University}
% %     % \country{China}
% }
% % \email{linmingfeng2023@email.szu.edu.cn}

% \author{Huayue Liang}
% \affiliation{
% %    \institution{Intel}
% %     % \country{China}
% }
% % \email{huayue.liang@intel.com}

\author{
  \begin{tabular}{c}
    Hongyi Guan\textsuperscript{*}, Yichuan Gao\textsuperscript{*}, Chenlu Miao\textsuperscript{$\ddag$}, Haoyang Wu, 
    Hang Zhu, Mingfeng Lin, Huayue Liang\\
  \end{tabular}
  \\[3mm]
}

\affiliation{
  \institution{Tsinghua University, Intel Labs China}
}
 \thanks{* Hongyi Guan and Yichuan Gao have contributed equally and are considered to be co-first authors.}
 \thanks{$\ddag$ Chenlu Miao is the corresponding author.}

\begin{abstract}
\label{abtract}

    Vector processors frequently suffer from inefficient memory accesses, particularly for const-stride and segment memory access patterns. While coalescing strided accesses is conceptually straightforward, implementing efficient data routing between memory and registers remains challenging. Conventional designs typically rely on high‐overhead crossbars that remap any byte in memory or registers to any position in registers or memory, leading to significant physical design issues. Meanwhile, segment operations requiring row-column transpositions force designers into an unfavorable trade-off: either employ element-by-element processing that severely compromises throughput, or implement large transposition buffers that significantly increase area and power consumption. These suboptimal approaches have created a fundamental gap in vector processor efficiency despite vectorization's theoretical advantages.

    In this paper, we present EARTH, a novel efficient vector memory access  architecture designed to overcome these challenges through shifting-based optimizations. 
    For const-stride accesses, EARTH integrates specialized shift networks for gathering and scattering strided elements. After coalescing multiple accesses into one request within the same cache line, data can be routed between memory and registers through the shifting network with minimal overhead. For segment operations, EARTH employs a shifted register bank that enables direct column‐wise access, eliminating the need for dedicated segment buffers while providing high‐performance, in‐place bulk transposition at acceptable overhead.
    
    We implemented the entire EARTH design on FPGA with Chisel HDL based on an open-source RISC-V vector unit Saturn.
    Our evaluation demonstrates that EARTH enhances performance for const-stride memory accesses proportionally to their prevalence in workloads, achieving 4x–8x speedups in benchmarks dominated by const-stride operations. The architecture also delivers area-efficient segment handling. Compared to conventional designs, EARTH reducing hardware area by 9\% and power consumption by 41\%. By optimizing these necessary memory access patterns, EARTH significantly advances both the performance and efficiency of vector processors.
\end{abstract}




\maketitle

\section{Introduction}
\label{sec:intro}

Vector processors offer flexible and efficient support for parallel computing across diverse fields such as finance, cryptography, signal processing, scientific computing, AI, etc. 
{They offer significant performance advantages for data‐intensive workloads by exploiting parallelism at scale. Although high‐performance systems often employ GPUs and specialized accelerators, these solutions can be costly, power‐hungry and inflexible, driving widespread interest in vector architectures—particularly in resource‐constrained environments—due to their balance of performance and efficiency.}

{Vector processors accommodate diverse workload characteristics through specialized memory access semantics, enabling efficient data movement and computation patterns.} In RISC-V Vector ISA \cite{rvvspec}, the memory access patterns are categorized into unit-stride, constant-stride, indexed, and segment operations. Among them, the first two stride patterns account for the vast majority of total memory accesses, while the other two contribute very little. Since data in memory is not always accessed in a sequential or aligned manner and there lacks effective mechanisms for data reorganizing, efficient memory load/store brings a huge challenge.

{
Our examination of state‐of‐the‐art open‐source vector designs reveals two principal bottlenecks, centered on a trade‐off between high performance and high overhead:
}

{
\begin{itemize}
    \item Inefficient handling of constant‐stride accesses: Many existing implementations issue multiple requests for the same cache line, limiting performance gains. Although coalescing these requests can mitigate some inefficiencies, naive methods rely on large crossbar interconnects for gather/scatter operations. Both approaches—issuing multiple requests or using crossbars—impose considerable overhead, introducing higher latency or more routing complexity.
    \item Suboptimal supporting of segment operations: Segment loads and stores require row-column transpositions that face an inherent trade-off. Element-by-element processing severely impacts throughput, while bulk transposition using large buffers consumes substantial area. Both approaches lead to suboptimal efficiency.
    % \item Offseting more reasonable trade-off for solving segment operations. Excessive overhead or insufficient performance for segment operations. Segment loads/stores require row‐column transpositions, which can be done element by element—severely degrading performance—or with large buffer‐based bulk transposition—incurring significant area overhead. Both approaches lead to suboptimal efficiency.
\end{itemize}
}

\vspace{-1mm}

To tackle above issues, we propose EARTH, 
{
 a novel vector memory access architecture that delivers performance on par with traditional high‐overhead methods while significantly reducing hardware costs.
}
This approach first-ever introduces shifting-based strategies into vector load/store unit designs, effectively addressing memory access issues for both strided and segment patterns while minimizing hardware resource requirements. The design of innovative data reorganization module (DROM) efficiently supports data gather and scatter through layered shifting networks, enabling systematic data reorganization across both memory and register operations.

Specifically, we adopt a new load/store data organization (LSDO) design for constant stride access patterns, enabling coalescing multiple accesses within the same cache line into a single memory request. We also leverage a novel row/column-accessible vector register file (RCVRF) to enable both row-wise and column-wise accesses, eliminating the need for dedicated segment buffers.

{
In summary, our paper makes following contributions:
\begin{itemize}
    \item We systematically analyze state-of-the-art open-source vector processors and pinpoint critical challenges in memory-access efficiency, including inadequate strided coalescing support and complex row-column transposition for segment operations (Section~\ref{sec:background} - Section~\ref{sec:overview}).
    \item 
        We propose EARTH, the first framework to incorporate shifting-based strategies for vector load/store operations, effectively handling both strided and segment patterns within a single design. EARTH addresses both strided and segment‐access inefficiencies through two key innovations: (1) a novel Load/Store Data Organization (LSDO) design that coalesces multiple accesses within the same cache line for constant-stride patterns. (2) A row/column-accessible vector register file (RCVRF) to streamline data movement and eliminate the need for dedicated segment buffers. (Section~\ref{sec:overview} - Section~\ref{sec:flow})
    \item We implemented EARTH using Chisel HDL \cite{jonathan2012chisel} and integrated it into Saturn, an open-source RISC-V vector unit that fully supports the RVV 1.0 application-profile specification. Our evaluation across various benchmarks demonstrates that EARTH achieves 4x–8x speedups on stride-intensive workloads while maintaining comparable performance on segment operations, all while reducing area overhead by 9\% and power consumption by 41\%. (Section~\ref{sec:evaluation})
\end{itemize}
}

\section{Background}
\label{sec:background}

In this section, we present essential background information to contextualize our work. We first introduce vector processing, then detail the memory access patterns specified in the RISC-V vector extension, and finally examine current vector processor designs and their approaches to memory access handling.

\subsection{Vectorization}

Data-intensive workloads have become pervasive across various fields, including finance, cryptography, signal processing, scientific computing, and AI \cite{gathu2024high, boemer2021intel}.
These workloads typically require processing vast amounts of independent data, posing significant challenges for traditional scalar processing architectures. To address this computational demand, various parallel processing techniques have emerged, with vectorization standing out as an effective approach. Vector processing, specifically through Single Instruction, Multiple Data (SIMD) architectures, offers a straightforward way to accelerate data-parallel operations \cite{flynn1972some, hennessy2017computer}. It allows a single instruction to operate on multiple data simultaneously, significantly improving throughput for applications with high data parallelism.{Vector processors have been central to high-performance computing ever since the Cray-1 supercomputer \cite{russell1978cray} demonstrated their effectiveness for scientific applications. }

{Compared to other parallel computing approaches—such as GPUs \cite{gpu} or Domain-Specific Architectures (DSAs) \cite{guo2023olive} -- vector processors offer notable advantages. Unlike GPUs, which impose complex thread management and synchronization overheads, they are generally more programmer-friendly and light-weight, facilitating easier integration into systems with stringent energy or area constraints \cite{hua2023edge}. Meanwhile, unlike DSAs, which often specialize in a narrow set of deep-learning operations, vector processors retain a flexible, general-purpose instruction set that accommodates varied computational kernels—from matrix arithmetic to cryptographic workloads.}

\begin{table}
    \centering
    \caption{Key Terminologies}
    \label{tab:rvv_terms}
    \vspace{-4mm}
    \small
    \begin{threeparttable}
    \begin{tabular}{p{1cm} p{6.5cm}}
        \toprule
        \textbf{Term} & \textbf{Description} \\
        \midrule
        VLEN & Number of bits available in a single vector register. \\
        
        ELEN & Maximum bit-width for individual vector elements. \\

        DLEN & Width of vector datapath \\

        MLEN & Width of vector memory interface \\

        VL & Vector length, representing the number of elements to be processed in a vector operation. \\

        EMUL & Effective Vector length multiplier, used to combine multiple vector registers into a single group. \\
                
        % SEW & Selected element width, typically set to 8, 16, 32, or 64 bits. \\
        
        EEW & Effective element width (8, 16, 32, or 64 bits). \\
        \bottomrule
    \end{tabular}
    \end{threeparttable}
    % \vspace{-5mm}
\end{table}

Traditional vector extensions, such as those found in x86 (such as AVX  \cite{intel_sdm, intel_avx512} and SSE \cite{intel_sdm}) and ARM (such as NEON \cite{arm_ddi0487}), often use fixed vector lengths, which limit their flexibility for different workloads. {By contrast, ARM’s Scalable Vector Extension (SVE) \cite{stephens2017arm}, inspired by the Cray-1 \cite{russell1978cray}, introduces variable-length vectors that adapt to workload requirements, improving performance across numerous application domains.}

The RISC-V Vector Extension (RVV) version 1.0 \cite{rvvspec} builds upon principles of flexibility and scalability, following the {variable-length approach}. Unlike traditional SIMD extensions with fixed vector length, RVV is designed to support variable-length vectors, making it suitable for a broad range of data processing tasks. This design enhances the versatility and efficiency of RISC-V, positioning it as a competitive, open-source option for diverse application scenarios. Table~\ref{tab:rvv_terms} provides definitions of the key terminologies that will be referenced throughout this paper.

\subsection{RVV Memory Access Patterns}

\begin{figure}
    \centering
    \includegraphics[width=\linewidth]{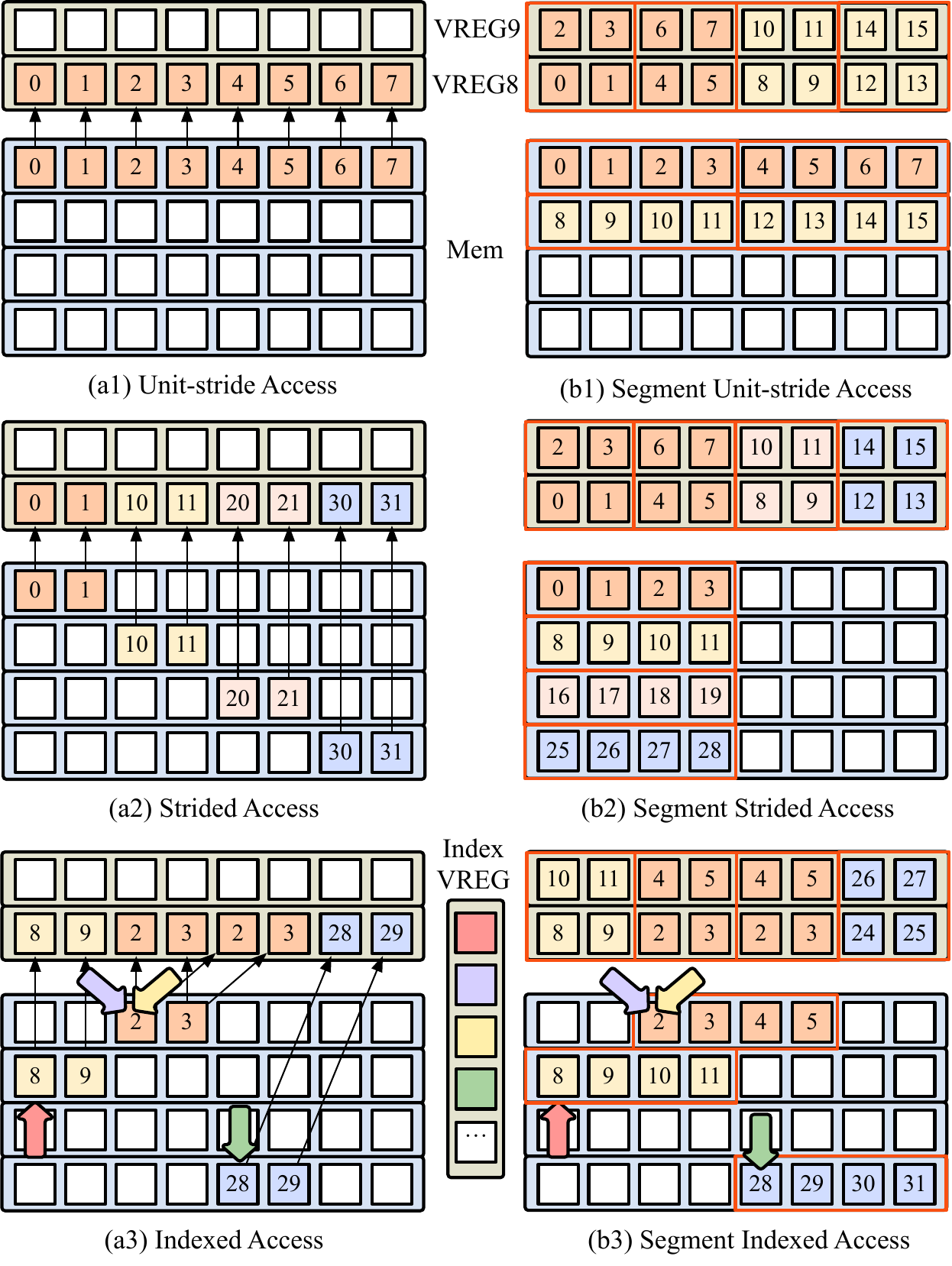}
    \vspace{-6mm}
    \caption{RVV Memory Access Patterns}
    \label{fig:mempattern}
    \vspace{-4mm}
\end{figure}

RVV supports diverse memory access patterns to efficiently handle varying data layouts. These patterns fall into four fundamental categories: unit-stride, strided, indexed, and segment operations. Figure~\ref{fig:mempattern} illustrates these memory access patterns, where each square block represents a single byte data. For this example, we consider a vector register with VLEN=64 bits (8 bytes), an EEW of 16 bits and a VL of 4 elements. Base is the starting memory address.

\subsubsection{Unit-stride Access.}
Unit-stride access is the most basic and efficient memory access pattern in RISC-V vector processing, where consecutive elements are accessed from contiguous memory locations. As shown in Figure~\ref{fig:mempattern}~(a1), vector register VREG8 loads eight bytes (labeled 0-7) sequentially from memory. For each element $i$, its memory address is calculated as:
\(
\text{Address}_i = \text{Base} + i \times \text{EEWB}
\)

\subsubsection{Strided Access.}
Strided access enables vector operations on non-contiguous memory locations separated by a constant stride. As shown in Figure~\ref{fig:mempattern}~(a2), for a stride of 10, vector register VREG8 loads elements from memory addresses with indices 0-1, 10-11, 20-21, and 30-31. Each element's memory address is calculated as:
\(
\text{Address}_i = \text{Base} + i \times \text{Stride}
\)

\subsubsection{Indexed Access.}
Indexed access, also known as scatter-gather, enables vector operations on arbitrary memory locations specified by an index vector. As shown in Figure~\ref{fig:mempattern}~(a3), the vector register loads four pairs of elements (8-9, 2-3, 2-3, 28-29) from memory locations determined by the index vector. Each element's memory address is calculated as:
\(
\text{Address}_i = \text{Base} + \text{Index}_i
\), 
where $\text{Index}_i$ is stored in a separate index vector register.

\subsubsection{Segment Access.}

Segment access is a sophisticated feature in RVV designed to efficiently handle Array-of-Structures (AoS) data layouts \cite{rvvspec}. This feature organizes vector register into logical segments, where each segment comprises elements from different vector registers. As illustrated in Figure~\ref{fig:mempattern}~(b1), consider the FIELD=2 case, where the two FIELD VREG8 and VREG9, each containing 4 elements. These registers are logically partitioned into segments, where each segment consists of two elements: one element from VREG8 and one from VREG9. When accessing an array of structures \texttt{arr} where each structure contains x and y of the same datatype:  \texttt{arr[0]} is written to the first segment: which means  \texttt{arr[0].x} is written to VREG8's first element,  \texttt{arr[0].y} to VREG9's first element and so forth.
RVV implements three variants of segment access: segment unit-stride, segment strided, and segment indexed. Each variant provides different memory addressing capabilities while maintaining the segment organization.

\noindent \textbf{Segment unit-stride access.} Segment unit-stride access operates by loading or storing data in consecutive memory locations in a structured way. As shown in Figure~\ref{fig:mempattern}~(b1), with FIELDS=2 and EEWB=2 bytes, each segment accesses four consecutive bytes. The first segment loads \texttt{memory[0-3]}, the second segment loads \texttt{memory[4-7]}, and so on. The elements are distributed across vector registers based on their positions within segments: \texttt{memory[0-1,\\4-5,8-9,12-13]} are written to VREG8, while \texttt{memory[2-3,6-7,\\10-11,14-15]} are written to VREG9. Each element's memory address can be computed using:
\(
\text{Address}_{i,j} = \text{Base} + i \times \text{FIELDS} \times \text{EEWB} + j \times \text{EEWB}
\), 
where $i$ is the segment index, $j$ is the field index within the segment.

\noindent \textbf{Segment strided access.}  Segment strided access loads from or stores to memory with a fixed stride between each segment. As shown in Figure~\ref{fig:mempattern}~(b2), with a stride of 8 between segments, the first segment loads \texttt{memory[0-3]}, the second segment loads \texttt{\\memory[8-11]}, followed by \texttt{memory[16-19]} and \texttt{memory[24-27]}. The elements are distributed across vector registers based on their positions within segments: \texttt{memory[0-1,8-9,16-17,24-25]} are written to VREG8, while \texttt{memory[2-3,10-11,18-19,26-27]} are written to VREG9. Each element's memory address can be computed using:
\(
\text{Address}_{i,j} = \text{Base} + i \times \text{Stride} + j \times \text{EEWB}
\)

\noindent \textbf{Segment indexed access.} Segment indexed access uses an index vector to determine the address of each segment. Figure~\ref{fig:mempattern}(b3) illustrates an example. Each element's memory address can be computed using:
\(
\text{Address}_{i,j} = \text{Base} + \text{Index}_i + j \times \text{EEWB}
\)

\begin{table}[t]
    \centering
    \caption{Comparison of Open Source RISC-V Vector Processors Designs}
    \label{tab:comparison}
    % \vspace{-2mm}
    \small
    \begin{threeparttable}
    \begin{tabular}{p{3cm} c c p{3cm}}
        \toprule
        \textbf{Design} & \textbf{UC\tnote{1}} & \textbf{SC\tnote{2}} & \textbf{Segment Support} \\
        \midrule
        Ara\tnote{3} \ ~\cite{perotti2022new} & \ding{51} & \ding{55} & Element-wise \\
        XiangShan\tnote{4} \ ~\cite{wang2024xiangshan} & \ding{51} & \ding{55} & Segment Buffer \\
        T1\tnote{5} \ ~\cite{website:t1} & \ding{51} & \ding{55} & Segment Buffer \\
        Saturn\tnote{6} \ ~\cite{website:saturn} & \ding{51} & \ding{55} & Segment Buffer \\
        EARTH & \ding{51} & \ding{51} & Buffer-free \\
        \bottomrule
    \end{tabular}
    \vspace{-6pt}
    \begin{tablenotes}
        \small
        \begin{multicols}{2}
            \item[1] UC: Unit-stride Coalescing
            \item[2] SC: Strided Coalescing
            \item[3] Ara commit: e6994c7
            \item[4] Xiangshan commit: f12520c
            \item[5] T1 commit: 13b2b16
            \item[6] Saturn commit: 49a04b9
        \end{multicols}
    \end{tablenotes}
    \end{threeparttable}
    \vspace{-6mm}
\end{table}

\subsection{{Challenges in Vector Memory Access Unit}}
{
Modern vector memory access units employ specialized methods to handle different memory access patterns. Table~\ref{tab:comparison} analyzes state-of-the-art open-source vector designs, revealing how they employ various techniques to handle diverse memory access patterns, yet face critical limitations. For \textit{unit-stride accesses}—which are contiguous—requests can be coalesced easily, thereby reducing memory transactions and efficiently utilizing memory bandwidth. Representative works \cite{cavalcante2019ara, wang2024xiangshan, website:t1, website:saturn} all implement this coalescing strategy for unit-stride operations.}

\textcolor{black}{
Current open-source designs for \textit{indexed access} employ no optimizations, relying on element-wise memory operations. Effective coalescing requires both address calculation for all elements and sophisticated logic to identify coalesceable accesses within cache lines. While AXI-Pack~\cite{zhang2023axi, zhang2024near} proposes an innovative near-memory computing approach that performs indexed element address computation directly in memory to avoid loading address indices into vector registers, their solution deviates from RVV indexed access semantics and lacks practical applicability in current systems.
}

\begin{figure}[ht]
    \centering
    \includegraphics[width=0.8\linewidth]{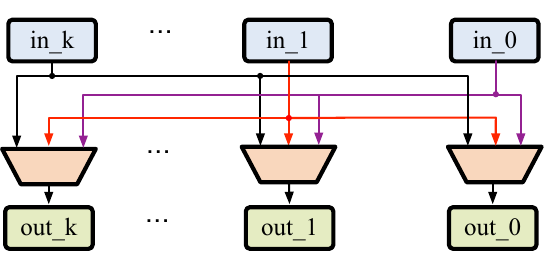}
    % \vspace{-2mm}
    \caption{\textcolor{black}{Crossbar Network for Byte-Level Remapping in Naive Strided Access Coalescing}}
    \label{fig:xbar}
    \vspace{-6mm}
\end{figure}

{
\textit{Strided accesses}—the second most common access pattern—pose a fundamental optimization challenge. Although extending coalescing to strided operations appears natural, naive approaches often incur high implementation costs. Replacing multiple smaller strided accesses with a single, larger coalesced request requires mapping any byte in source to any byte in destination—which is a nontrivial task. Achieving this fine-grained mapping typically demands crossbars between memory and vector registers, incurring significant area and power overhead while also complicating physical design, as illustrated in Figure~\ref{fig:xbar}. As VLEN or MLEN grows, crossbar complexity proliferates, booming both cost and complexity. Consequently, naive coalescing methods fail to deliver the anticipated performance benefits within realistic design constraints. AXI-Pack \cite{zhang2023axi} proposes a strategy to accelerate strided-memory access by modifying the AXI protocol, merging multiple strided requests into fewer, larger transactions—thus reducing transaction overhead at the cost of requiring custom extensions to the memory subsystem and interconnect.}

\begin{figure}[t]
    \centering
    \includegraphics[width=0.95\linewidth]{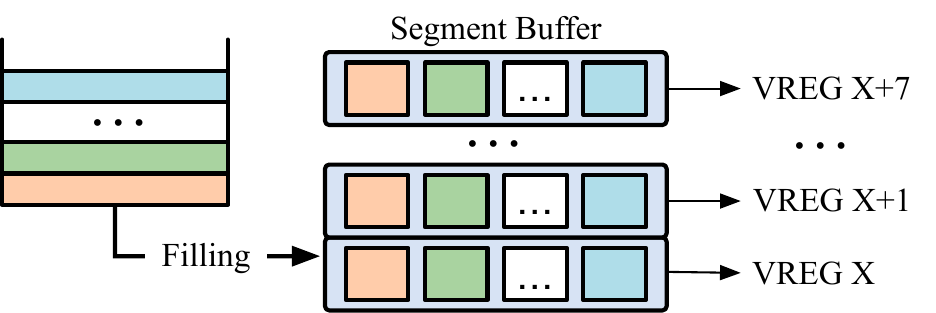}
    % \vspace{-2mm}
    \caption{Segment Buffer}
    \label{fig:segbuffer}
    \vspace{-4mm}
\end{figure}

\textcolor{black}{
For \textit{segment accesses}, current designs generally fall into one of two categories: an element-wise approach or a segment buffer approach. In the element-wise approach, as adopted by Ara \cite{cavalcante2019ara}, segment instructions are decomposed into individual elements. This simplifies data transposition but can severely increase memory access overhead. In contrast, the segment buffer approach uses dedicated buffers to coalesce requests within segments, reducing the number of memory transactions. However, it introduces considerable hardware overhead for row-column transposition \cite{website:t1, wang2024xiangshan, website:saturn}. Figure~\ref{fig:segbuffer} shows a classic segment buffer design and its processing flow: the buffer accumulates source data column-by-column until forming complete rows—at which point it writes the data to the destination in a manner compatible with row-major organization.
}

\section{Overview}
\label{sec:overview}

In this section, we present EARTH, a novel architecture that optimizes vector memory accesses while keeping hardware costs low. Vector memory access patterns remain a key performance bottleneck in modern processors. 
While existing open-source vector designs handle unit-stride memory operations well, they struggle with constant-stride patterns. Current approaches also rely on dual segment buffers that use substantial chip area without delivering matching performance gains. 
EARTH solves these problems through three key innovations. First, at the heart of EARTH lies the innovative data reorganization module (DROM). DROM efficiently supports data gather and scatter through layered shifting networks, enabling systematic data reorganization across both memory and register operations. Second, building upon DROM, the Load/Store Data Organization Module (LSDO) organizes data for strided access patterns, enabling multiple memory requests within aligned MLEN regions to be combined into single transactions. Third, the Row/Column-accessible Vector Register File (RCVRF), also leveraging DROM, uses its Shifted VRF design to support dual-access patterns without needing segment buffers to support segment operations, maintaining high performance while reducing hardware complexity. \textcolor{black}{We integrate EARTH into Saturn~\cite{website:saturn}, a general RISC-V vector implementation. For simplicity, we refer to the integrated system as EARTH throughout the rest of this work.}

\subsection{Motivation}
\label{subsec:motivation}

Memory access significantly impact vector processor performance, often creating a severe bottleneck in achieving peak efficiency.  Current Vector LSUs, though effective at coalescing unit-stride operations, fail to optimize strided access patterns, leaving substantial performance potential untapped through missed coalescing opportunities. Additionally, conventional designs' reliance on segment buffers for segment operations introduces excessive area overhead and compromises resource efficiency. These critical limitations in both performance and efficiency underscore the need for a fundamentally new approach to handle vector memory access.

\noindent\textbf{Limited Hardware Support for Strided Access Coalescing.}
Strided access patterns, despite being prevalent across diverse benchmarks, suffer from inefficient hardware support that fails to exploit available performance opportunities. This limitation primarily stems from a fundamental challenge: the absence of efficient data reorganization mechanisms to handle load/store operations. For loads, the hardware lacks support to extract strided elements from coalesced response, while for stores, it cannot efficiently scatter register data to appropriate memory positions. Current designs resort to naive element-wise decomposition, generating redundant memory requests to the same aligned MLEN region. Consider a concrete example: a vector load instruction requests 32 1-byte elements with 2-byte stride (MLEN = 64 bytes). Although all elements could potentially reside within a single 64-byte cache line, the operation triggers 32 separate cache accesses. This inefficiency results in two critical performance bottlenecks: (1) increased latency from serialized cache accesses, and (2) wasted memory bandwidth due to redundant requests to the same cache line.

\noindent\textbf{Inefficient Hardware Resources for Segmented Access.}
Segment operations present a challenge of efficiently managing both memory operations and data transposition. These operations, which handle data transformation between row and column formats, face fundamental implementation challenges due to vector register files' inherent limitation to row-wise access. Current approaches to supporting segment accesses involve significant trade-offs. The element-wise method decomposes segment instructions into individual elements, simplifying transposition but incurring substantial memory access overhead \cite{cavalcante2019ara, perotti2022new}. Common designs \cite{website:t1, wang2024xiangshan, website:saturn} employ dedicated segment buffers to coalesce memory requests within segments, but introduce considerable hardware overhead for row-column transposition.
To illustrate these trade-offs, let's consider segment load operations under two current approaches. The element-wise approach processes data sequentially, requiring $\text{FIELD} \times \text{VL}$ discrete memory accesses per segment instruction --- a clear performance bottleneck. The prevalent buffer-based approach implements dedicated segment buffers for data reorganization. While more efficient than element-wise processing, this approach demands substantial hardware resources: the RISC-V vector specification's support for up to eight vector registers in segment operations necessitates dual segment buffers, each sized at $8\times \text{MLEN}$, for separate load and store requests. This significant area overhead is particularly questionable, especially given that segment instructions are not commonly used in practical applications.

\subsection{Methodology}

EARTH introduces novel shifting-based strategies that simultaneously optimize vector memory access performance and minimize hardware complexity. As shown in Table~\ref{tab:comparison}, EARTH achieves both unit-stride and strided memory access coalescing, while supporting segment operations without dedicated buffers. Our approach introduces three key architectural innovations that address fundamental limitations in contemporary vector architectures:

\noindent \textbf{Shift Networks Enable Advanced Data Reorganization.} 
We propose a novel DROM to systematically handle efficient data gathering and scattering. At its core, DROM incorporates shift networks, including Scatter Shift Network (SSN) and Gather Shift Network (GSN). DROM serves as a foundational component in both LSDO and RCVRF.

\noindent \textbf{LSDO Facilitates Coalesced Strided Access Data Handling.} LSDO is designed to handle the organization of strided access data by employing a Reverser and DROM. By leveraging LSDO, our design coalesces multiple accesses within the same aligned MLEN region into a single memory request while maintaining proper data arrangement for strided operations. This reduces memory bandwidth consumption and enhances overall performance.

\noindent \textbf{RCVRF Supports Segment Access Without Segment Buffers.}
EARTH introduces an innovative RCVRF composed of Shifted VRF and DROM, which natively supports both row-wise and column-wise access. This dual-access capability eliminates the need for dedicated segment buffers, substantially reducing hardware overhead while fully supporting segment operations.

\subsubsection{Shift Networks Enable Advanced Data Reorganization}

DROM serves as the central component of EARTH's data handling infrastructure, with its Shift Networks -- comprising SSN and GSN -- forming the cornerstone of data reorganization capabilities. DROM architecture integrates a Shift Count Generation Module (SCG) that dynamically controls the shift operations by generating appropriate shift counts for the networks.

DROM addresses two fundamental data reorganization challenges: scattering, which transforms stride-separated elements into sequential data, and gathering, which reorganizes contiguous data into stride-separated positions. To efficiently handle these operations, SSN and GSN implement a layered shift network architecture where each level enables power-of-2 shifts, allowing data elements to progressively reach their target positions. This hierarchical design ensures both flexibility and scalability in reorganization tasks.

\subsubsection{LSDO Facilitates Coalesced Strided Access Data Handling}
\label{subsubsec:shift_net}

To address the challenge of data organization in coalesced strided access, we propose LSDO. LSDO integrates DROM and Reverse module to organize strided access data. This architecture enables efficient handling of diverse stride patterns, supporting both positive and negative strides, as well as power-of-2 and non-power-of-2 data reorganization. For strided load operations, LSDO first processes negative strides through the Reverse module before passing the data to DROM for reorganization, ultimately producing the required output data pattern. For store operations, the data flow follows the symmetrical path.

 % To address the challenge of data organization stems from coalescing strided access, we propose the Load Store Data Organization (LSDO) module to do organization, which integrates a Data Reorganization Module (DRM) and a Reverse module. This architecture enables efficient handling of diverse stride patterns, supporting both positive and negative strides, as well as power-of-2 and non-power-of-2 data reorganization. For strided load operations, the LSDO first processes negative strides through the Reverse module before passing the data to the DRM for reorganization, ultimately producing the required output data pattern. For store operations, the data flow follows the reverse path.

\subsubsection{RCVRF Supports Efficient In-place Segment Access}
\label{subsubsec:shift_register}

EARTH addresses segment operations challenges through RCVRF. RCVRF integrates shifted VRF and DROM to achieve efficient data handling. The shifted VRF is partitioned into eight ELEN-bit banks, where corresponding elements from eight consecutive registers are distributed across banks, enabling parallel column access. While this VRF structure supports parallel access, it requires DROM to handle necessary data reorganization for column operations. For column access, DROM gathers data during reads (e.g., collecting the first byte from registers V0-V7 into contiguous data) and scatters data during writes. Both row and column access patterns utilize a block shifter for proper data alignment.

\begin{figure}[htb]
    % \vspace{-3mm}
    \centering
    \includegraphics[width=\linewidth]{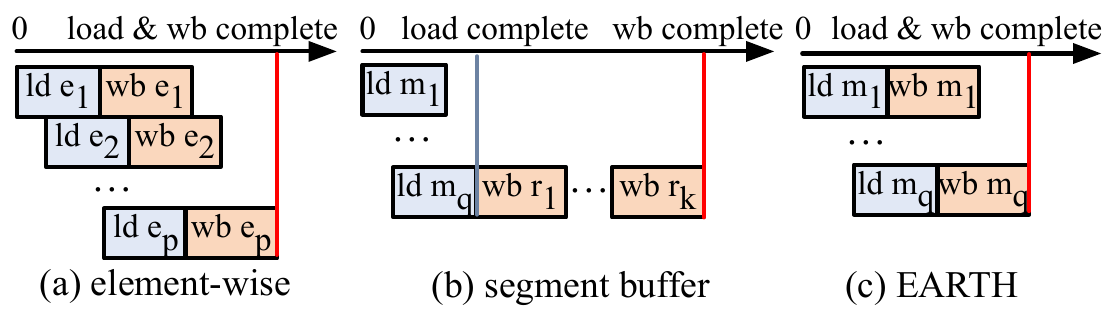}
    \vspace{-4mm}
    \caption{Timeline of methods to support segment intructions}
    \label{fig:timeline}
    % \vspace{-4mm}
\end{figure}

Figure~\ref{fig:timeline} illustrates the efficiency gains of EARTH compared to existing approaches. Consider a segment access with $p$ elements ($p=\text{FIELDS}\times\text{VL}$), where elements within the same segment reside in the MLEN region. The access involves $q$ segments, resulting in $q$ memory requests, with each segment distributed across $k$ vector registers. The element-wise approach (Figure~\ref{fig:timeline}(a)) implements a simple but inefficient pipeline of loading (ld $e_i$) and writing back (wb $e_i$) for individual elements. 
The traditional segment buffer approach (Figure~\ref{fig:timeline}(b)) reduces memory requests to $q$ but introduces a rigid two-phase operation: bulk loading into segment buffers  (ld $m_i$)) followed by sequential row-wise writebacks (wb $r_i$) to vector registers. In contrast, EARTH's shifted register approach (Figure~\ref{fig:timeline}(c)) achieves both reduced memory requests and sustained pipeline efficiency by enabling immediate writeback (wb $m_i$) following each memory load (ld $m_i$).

\section{EARTH Architecture}
\label{sec:architecture}

\begin{figure*}
\centering
\includegraphics[width=\linewidth]{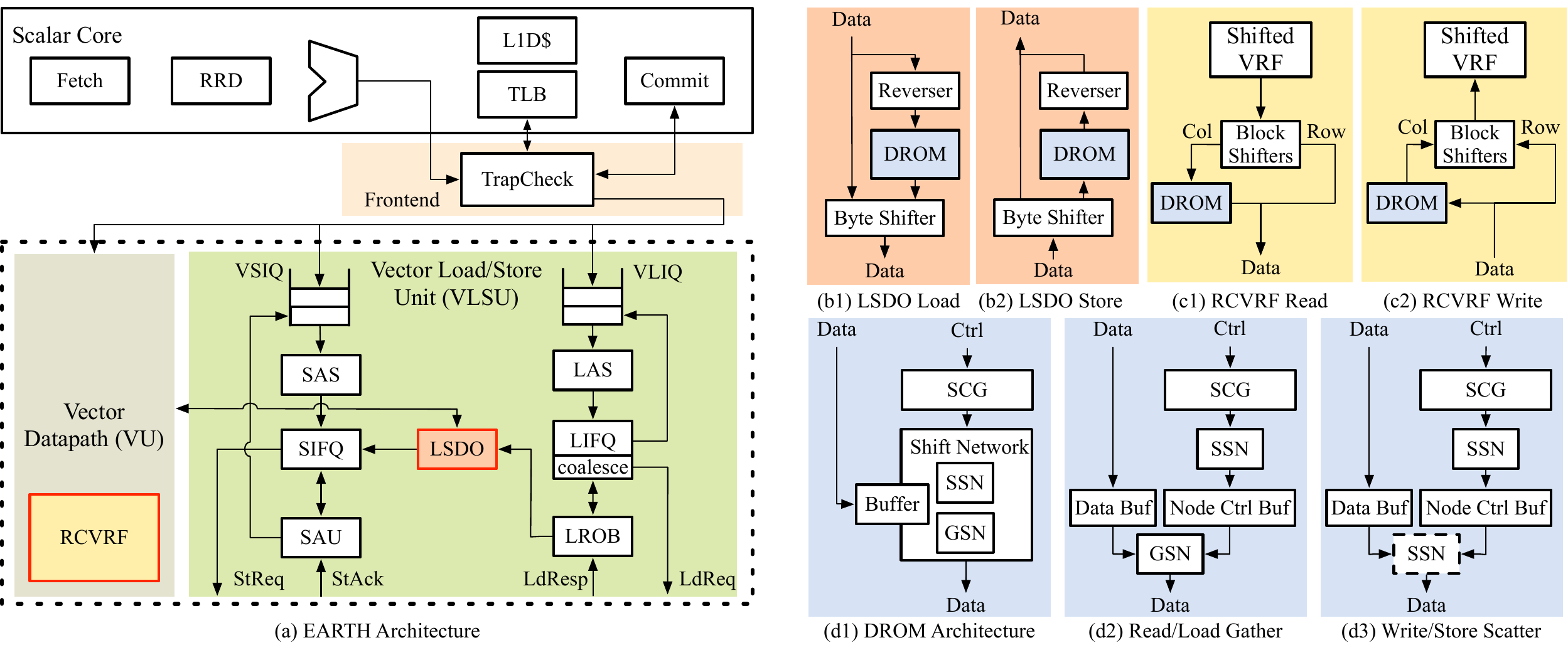}
\vspace{-6mm}
\caption{EARTH Architecture Overview}
\label{fig:earth}
\end{figure*}

\textcolor{black}{
EARTH (together with Saturn) consists of three primary modules, as depicted in Figure~\ref{fig:earth}: the Vector Frontend Unit (VFU) for trap checks of vector operations; the Vector Datapath Unit (VU) for executing arithmetic operations, with vector registers residing within it; and the Vector Load/Store Unit (VLSU) for managing memory operations. The architecture of EARTH incorporates an efficient VLSU and RCVRF to enable high-performance data handling. The RCVRF features a shifted register bank design that directly supports both row-wise and column-wise register accesses. The VLSU includes several modules to effectively manage memory operations, with the Load/Store Data Organizer (LSDO) being central to its efficiency. Additional modules comprise the Load/Store Address Sequencer (LAS/SAS), which splits memory accesses into operations based on element width or alignment with memory width boundaries, and the Load/Store In-Flight Queue (LIFQ/SIFQ) maintains the ordering of memory operations, working in conjunction with the Load Reordering Buffer (LROB) and the Store Acknowledgement Unit (SAU) to manage out-of-order arrival data and acknowledgments.
A key component shared by both the LSDO and the RCVRF is the Data Reorganization Module (DROM), as shown in Figure~\ref{fig:earth}~(d1). The DROM consists of two essential parts: the Shift Networks, including GSN and SSN, and the SCG. These components play a pivotal role in optimizing data reorganization within the LSDO of the VLSU and the RCVRF. Specifically, the LSDO efficiently handles data reorganization for strided accesses, while the RCVRF, utilizing its shifted register bank design and the DROM, facilitates direct row-wise and column-wise accesses without requiring dedicated segment buffers. The detailed design of EARTH will be explored in subsequent sections. Furthermore, Section~\ref{sec:flow} will elaborate on the processing flows for various memory access patterns, demonstrating the practical impact of this architecture.
}

\subsection{Shift Networks}

\begin{figure}
    \centering
    \includegraphics[width=\linewidth]{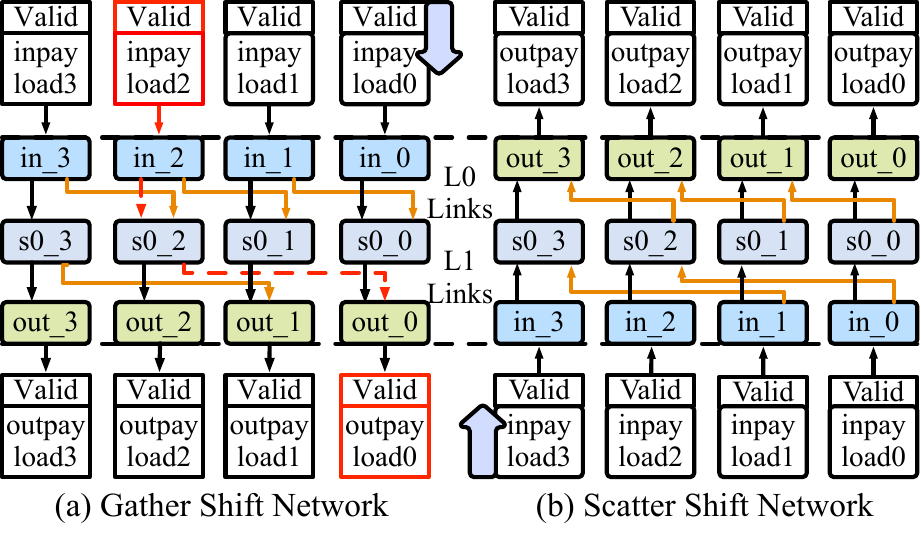}
    \vspace{-6mm}
    \caption{Shift Network Architecture}
    \label{fig:shiftnet}
    \vspace{-4mm}
\end{figure}

EARTH employs two types of shift networks: GSN and SSN, as shown in Figure~\ref{fig:shiftnet}. These networks are designed with opposing data flow directions to ensure conflict-free operations -- GSN facilitates top-down flow, while SSN implements bottom-up flow. Given that SSN mirrors GSN's functionality with reversed logic, we will focus on GSN's design.

\subsubsection{Shift Operation}
GSN performs routing operations on vectors of size \(n\), where each element contains both valid and payload fields: \(\text{vec}(n, \{ \text{valid}, \text{payload} \})\).
For each input element at column \(i\), GSN routes it to an column \(j\) through a series of right shift operations. The required shift amount,  \(\text{shiftCnt} = |i - j|\) 
is decomposed into its binary representation:
\[
\text{shiftCnt} = b_{L-1} \cdots b_0, \text{ where } L=\log_2(n)
\]
This binary decomposition enables an efficient layered implementation, where each layer \(l\) performs a right shift of \(2^l\) positions when its corresponding bit \(b_l\) is 1, and no shift when \(b_l\) is 0.

\subsubsection{Network Organization}

GSN implements shift operations through a hierarchical network composed of specialized nodes interconnected by two types of links across multiple layers. As illustrated in Figure~\ref{fig:shiftnet}~(a), the network processes vector elements and their validity signals through this Node-Link structure to achieve the desired shift operations.

\paragraph{Nodes}

\begin{figure}[t]
    \centering
    \includegraphics[width=\linewidth]{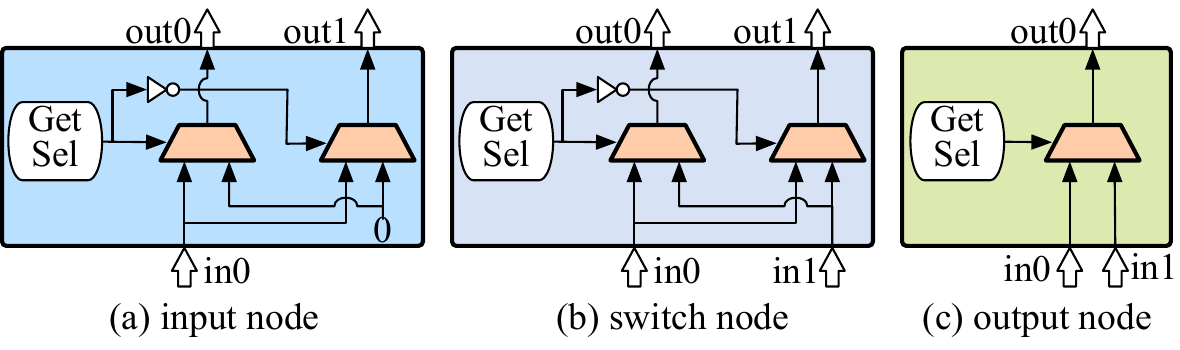}
    \vspace{-6mm}
    \caption{Three types of nodes in the network architecture.}
    \label{fig:node}
    \vspace{-4mm}
\end{figure}

The network architecture incorporates three specialized node types, depicted in Figure~\ref{fig:node}:
\begin{itemize}
\item \textbf{Input Nodes} (Figure~\ref{fig:node}~(a)): Located at Node Layer 0 in GSN, these nodes process incoming elements containing payload and validity signals. Based on their selection signals, each node routes valid inputs to either \(out_0\) or \(out_1\), corresponding to straight and diagonal links respectively.

\item \textbf{Switch Nodes} (Figure~\ref{fig:node}~(b)): Positioned in intermediate layers, these nodes implement the core switching logic. Each switch node processes two inputs (\(in_0, in_1\)) and, controlled by its selection signal, either maintains or exchanges their order to produce two outputs (\(out_0, out_1\)).

\item \textbf{Output Nodes} (Figure~\ref{fig:node}~(c)): Situated in the final layer in GSN, these nodes receive two inputs where exactly one is valid, and forward only the valid input to their output.
\end{itemize}

The selection signals for all nodes are derived from either the shift count information embedded in the input data stream or external control modules based on the required shift configuration.

\paragraph{Links}
Each link layer \(l\) between adjacent node layers \(l\) and \(l+1\) employs two distinct connection types:
\begin{itemize}
\item \textbf{Straight Links}: Establish direct vertical connections between corresponding nodes in adjacent layers (e.g., \(in\_2 \rightarrow s0\_2\)), preserving column positions.

\item \textbf{Diagonal Links}: Create non-circular shifted connections, routing data \(2^l\) positions rightward to the next layer (e.g., \(in\_2 \rightarrow s0\_1\)). Unlike circular shift networks, diagonal links do not wrap around to create circular connections.
\end{itemize}

\subsubsection{Example Walkthrough}
Figure~\ref{fig:shiftnet} demonstrates GSN's routing capability through a representative example, highlighted by a red dashed path. Consider routing an input from position 2 to position 0, requiring a shift count \(\text{shiftCnt} = |2-0| = 2 = (10)_2\), where the binary representation indicates \(b_1 = 1\) and \(b_0 = 0\). The routing process proceeds through three node layers:

\noindent \textbf{Node Layer 0}: The payload enters at input node \(in\_2\). Since \(b_0 = 0\), the input node routes the data through its straight output. The payload traverses the straight link in Link Layer 0 to reach switch node \(s0\_2\) in Node Layer 1.
    
\noindent  \textbf{Node Layer 1}: At this layer, \(b_1 = 1\) triggers an exchange operation. The switch node routes the payload through its diagonal link in Link Layer 1, directing it to output node \(out\_0\) in Node Layer 2.
    
\noindent \textbf{Node Layer 2}: The payload arrives at output node \(out\_0\), which forwards it to the final output position, completing the two-position right shift operation.

\subsubsection{Conflict-Free Property of the Shift Network}
\label{subsubsec:conflict-free}
SSN and GSN are designed to be conflict-free, ensuring efficient data routing without path interference. This property is guaranteed by two fundamental characteristics: \textit{order-preserving} and \textit{separation-preserving}.

\noindent\textbf{Order-preserving Property:} 
For \(k \geq 2\) valid inputs with positions \(pos_{in_1}, pos_{in_2}, \ldots, pos_{in_k}\) where:
\(
pos_{in_1} \leq pos_{in_2} \leq \ldots \leq pos_{in_k}
\)
Their corresponding output positions maintain the same order:
\(
pos_{out_1} \leq pos_{out_2} \leq \ldots \leq pos_{out_k}
\)

\noindent\textbf{Separation-preserving Property:}
The network maintains specific separation rules based on operation type:
\begin{itemize}
   \item \textbf{Scatter}: Preserves or increases element separation:
   \[
   |pos_{out_x} - pos_{out_y}| \geq |pos_{in_x} - pos_{in_y}|, \quad \forall x, y \in \{1, \ldots, k\}
   \]
   
   \item \textbf{Gather}: Preserves or decreases element separation:
   \[
   |pos_{out_x} - pos_{out_y}| \leq |pos_{in_x} - pos_{in_y}|, \quad \forall x, y \in \{1, \ldots, k\}
   \]
\end{itemize}

These properties ensure no path conflicts occur in the network. We prove this for GSN through contradiction (the same logic applies to SSN):

\begin{proof}[Proof of Conflict-Free Property]
Suppose two inputs \(in_a\) and \(in_b\) meet at node \((l,k)\) (Node Layer \(l\), column \(k\)). We show this leads to a contradiction:

\noindent 1) After meeting at layer \(l\), the paths must separate in some layer \(t > l\) due to different output columns, where one path shifts right by \(2^t\) and the other stays straight.

\noindent 2) For a GSN, the output separation must not exceed the input separation:
\[
|pos_{out_b} - pos_{out_a}| \geq 2^t \implies |pos_{in_b} - pos_{in_a}| \geq 2^t
\]

\noindent 3) However, the maximum possible input separation for paths meeting at node \((l,k)\) is:
\[
|pos_{in_b} - pos_{in_a}| \leq 2^l - 1 < 2^t
\]

This contradicts step 2, proving that two paths cannot meet at any intermediate node without violating the separation property. Therefore, the network is conflict-free.
\end{proof}

\subsubsection{\textcolor{black}{Physical design complexity}}
\label{subsubsec:phys_design}

\textcolor{black}{ 
We structured the GSN and SSN layers to allow only vertical or equidistant unidirectional data movement as shown in Figure~\ref{fig:shiftnet}, which allows us to easily complete physical design in the backend process of ASIC, while occupying only a minimal number of metal layers.
}

\subsection{Shift Count Generation}

SCG computes the required shift distance for each vector element. For a strided vector access with stride, EEWB and offset, the shift count is calculated as:
\[
  \text{shiftCnt}_i = (\text{stride} - \text{EEWB}) \times \lfloor \frac{i}{EEWB} \rfloor + \text{offset}
\]
where $i$ represents the destination position in scatter operations or source position in gather operations.

\begin{figure}[h]
  \centering
  % \vspace{-2mm}
  \includegraphics[width=\linewidth]{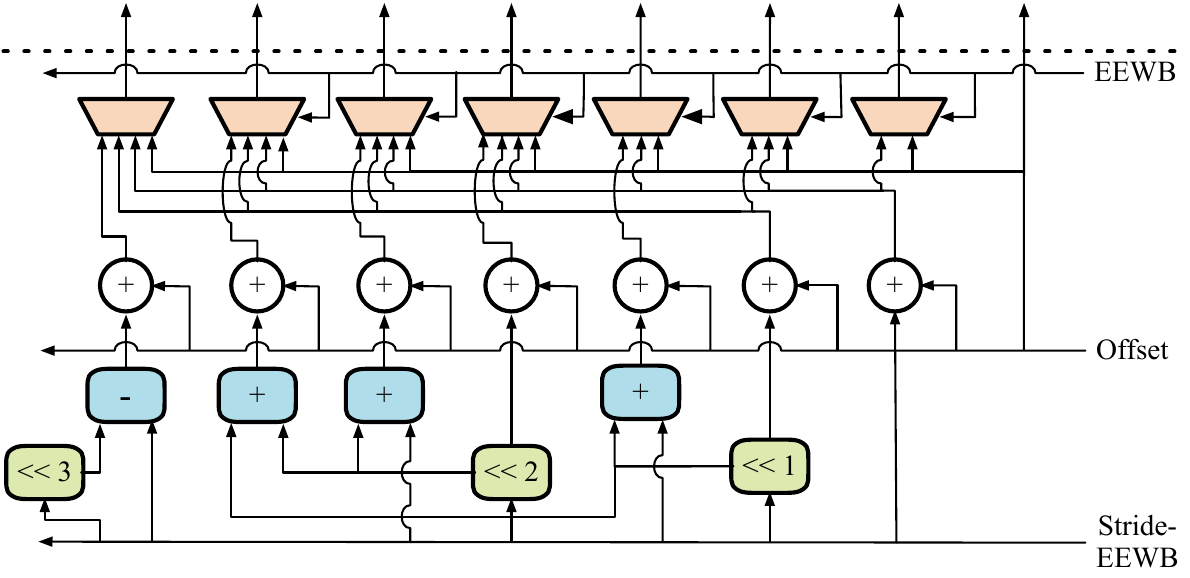}
  \vspace{-7mm}
  \caption{Shift Count Generation}
  \label{fig:shiftcode}
  % \vspace{-4mm}
\end{figure}

As shown in Figure~\ref{fig:shiftcode}, SCG generates these shift counts through three efficient steps: 1) Calculate \((\text{stride} - \text{EEWB}) \times i\) using shift and add/sub operations. 2) Add \(\text{offset}\) to generate position values. 3) Select final shift counts based on EEWB using multiplexers

For example, consider a strided load with \(\text{stride} = 4\), \(\text{EEWB} = 2\) and \(\text{offset} = 2\). This operation maps:
\begin{itemize}
  \item Input bytes [2,3] \(\rightarrow\) Output bytes [0,1]: shift right by 2
  \item Input bytes [6,7] \(\rightarrow\) Output bytes [2,3]: shift right by 4
  \item Input bytes [10,11] \(\rightarrow\) Output bytes [4,5]: shift right by 6
  \item Input bytes [14,15] \(\rightarrow\) Output bytes [6,7]: shift right by 8
\end{itemize}

\subsection{Data ReOrganization Module}
Shift Networks (SSN and GSN) and SCG constitute the core DROM in EARTH. As shown in Figure~\ref{fig:earth}~(d1)-(d3), each DROM comprises an SSN, GSN, SCG, and associated buffers, supporting both gather and scatter operations through distinct data paths. For read/load (gather) operations, DROM processes data and control signals as follows:

\begin{itemize}
\item Control signals (stride, EEWB, offset, etc.) feed into SCG to calculate shift counts that map input data elements to their correct output positions.
\item SSN processes shift counts to identify valid data elements and generate corresponding GSN node control signals.
\item Node control signals and input data are buffered in Node Ctrl Buffer and Data Buffer respectively.
\item GSN combines buffered data and control signals to produce gathered (sequential) data.
\end{itemize}

The write/store (scatter) operation uses a similar process, with SSN serving dual roles: first generating node control signals, then performing data scattering based on the buffered control signals.

\subsection{Load/Store Data Organization}
DROM serves as a key component within LSDO pipeline. As shown in Figure~\ref{fig:earth}~(b1)-(b2), LSDO comprises Reverser, DROM  and Byte Shifter. The Reverser handles negative stride operations, while the Byte Shifter performs alignment of data to specific offset. During load operations (Figure~\ref{fig:earth}~(b1)), input data flows from top to down through the pipeline. For non-strided access, data can bypass both the Reverser and DROM, proceeding directly to the Byte Shifter for final alignment. For strided access, data passes through the Reverser when stride is negative, then through DROM for gathering operations, and finally through the Byte Shifter for offset adjustment.
Store operations (Figure~\ref{fig:earth}~(b2)) utilize the same components but in reverse flow, with data moving from bottom to up through the Byte Shifter, DROM and Reverser.

\subsection{Row/Column-accessible Vector Register File}

EARTH introduces RCVRF, a novel design that enables bidirectional (row-wise and column-wise) vector data access while eliminating the overhead traditionally associated with segment buffers. The RCVRF architecture comprises three key components: Block Circular Shifters, DROM and Shifted VRF. Unlike the barber’s pole VRF design introduced by Ara \cite{cavalcante2019ara}, which does not support column-wise access due to its lack of a data reorganization mechanism, RCVRF overcomes these limitations through innovative design.

\subsubsection{Shifted Vector Register Organization}  
RCVRF partitions the vector register file into \textit{nBanks} = 8 banks, corresponding to the maximum number of vector registers accessible by a single instruction. Each bank has a width of ELEN bits (typically 64 bits), with each unit referred to as an ELEN Block. The number of rows per bank, denoted as \textit{nRows}, is given by \( nRows = \text{VLEN} \times 32 / (\text{ELEN} \times nBanks) \). The architecture employs a circular-shifted mapping scheme. The mapping function \( f \) is formally defined as:
\begin{equation*}\vspace{-2mm}
(\text{VREG}_i, \text{ELEN\_Block}_j) \xrightarrow{f} (Bank_k, Row_r)
\end{equation*}
where:
\begin{align*}
k &= (i + j) \bmod nBanks \\
r &= (\lfloor \frac{i}{nBanks} \rfloor \times \frac{\text{VLEN}}{\text{ELEN}} + i \bmod nBanks) \bmod nRows
\end{align*}

This mapping establishes a diagonal pattern with two essential properties: 
First, consecutive elements within a vector register map to consecutive banks, enabling efficient single-register access. Second, corresponding elements across different registers distribute across distinct banks, facilitating parallel access.

\begin{figure}
    \centering
    \includegraphics[width=\linewidth]{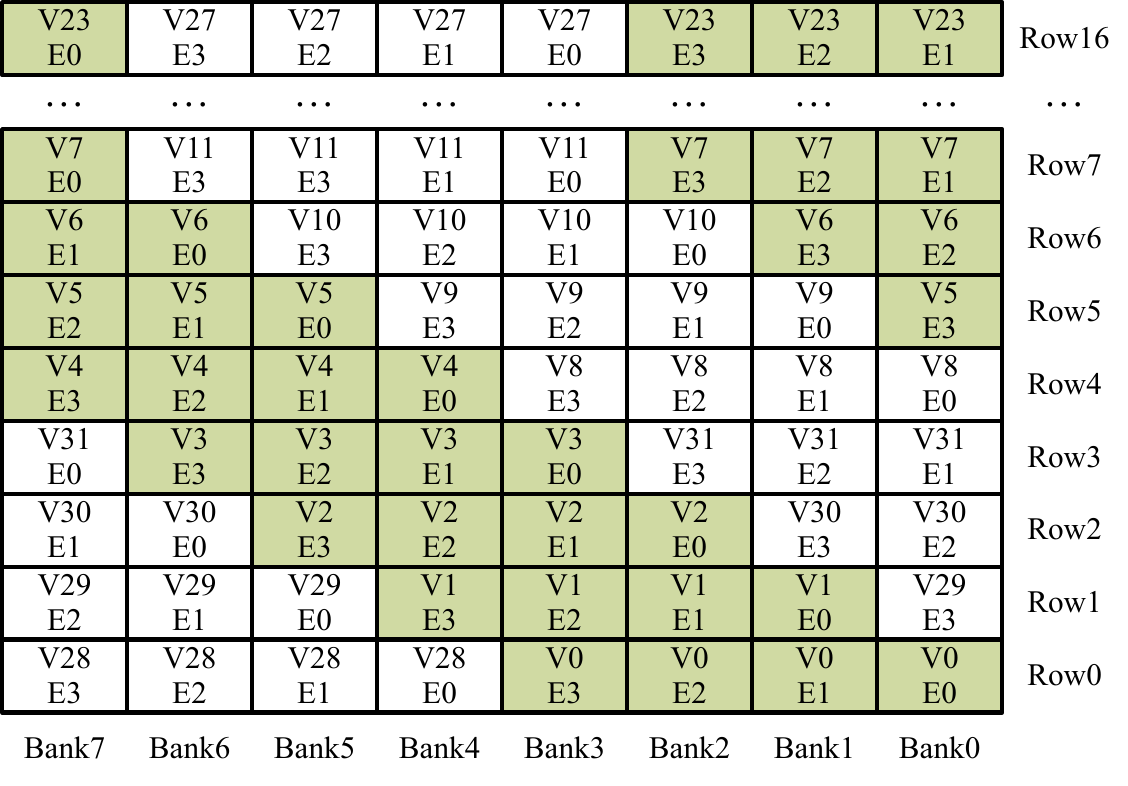}
    \vspace{-4mm}
    \caption{Shifted VRF When VLEN=256, ELEN=64}
    \label{fig:shiftedvreg}
    \vspace{-4mm}
\end{figure}

For VLEN=256, ELEN=256, as illustrated in Figure~\ref{fig:shiftedvreg}, it yields:
\begin{itemize}
\item VREG\textsubscript{0}: ELEN Blocks in Row\textsubscript{0}'s Bank\textsubscript{0}, Bank\textsubscript{1}, Bank\textsubscript{2}, ...
\item VREG\textsubscript{1}: ELEN Blocks in Row\textsubscript{1}'s Bank\textsubscript{1}, Bank\textsubscript{2}, Bank\textsubscript{3}, ...
\item ...
\item VREG\textsubscript{7}: ELEN Blocks in Row\textsubscript{7}'s Bank\textsubscript{7}, Bank\textsubscript{0}, Bank\textsubscript{1}, ...
\end{itemize}

\subsubsection{Access Mechanisms}
Figure~\ref{fig:earth}~(c1)-(c2) illustrates the data access flow, using a read process as an example.

\noindent\textbf{Row-wise Access:} For row-wise access, also means single register access, the Block shifter performs circular shifts, i.e. shifting vreg's ELEN\_Block 0 to position 0 for reads.

\noindent\textbf{Column-wise Access:} Column-wise access involves reading or writing the same element across vector registers. For example, when reading the first bytes from V0E1 through V7E1, all banks are accessed in parallel, retrieving the required data (V0E1, V1E1, ..., V7E1). These elements are initially read in the order (V6E1, ..., V0E1, V7E1). The Block Shifter then performs circular shifts to align the data in the order (V7E1, V6E1, ..., V0E1). The aligned data is subsequently processed by DROM, which utilizes the SCG to compute the required shift count. This shift count is based on a const stride value of \( \text{EMUL} \times \text{ELEN}/8\). Following DROM's read process, target bytes are consolidated into sequential output (V7E1's byte0, V6E1's byte0, ..., V0E1's byte0).

\section{EARTH Flow}
\label{sec:flow}

\subsection{Strided Access}
For strided load operations, instructions from the VLIQ head are directed to LAS. LAS splits instructions based on stride and MLEN, optimizing memory access by coalescing the maximum number of stride elements within a single aligned MLEN memory region. For each split operation, LAS allocates an entry in LIFQ to store control information and issues these requests to L2 sequentially. Memory responses from L2 are processed in order. While responses may arrive out of order and are temporarily stored in LROB, only ordered responses flow to LSDO for processing. LSDO orchestrates data reorganization guided by control signals from the corresponding LIFQ entry. Within LSDO, SCG and SSN generate precise control signals that direct GSN's data gathering process. The gathered data undergoes byte-level shifting for proper alignment. Finally, LSDO writes the processed results to RCVRF in a row-wise manner, completing the strided access operation.

For strided store operations, SAS generates split mops and allocates corresponding entries in SIFQ. SIFQ reads data from RCVRF in a row-wise manner, directing this vector register data to LSDO for data scattering. After data reorganization, SIFQ issues strided store requests to L2. Each SIFQ entry remains active until its corresponding store acknowledgment returns from L2, at which point the entry can be dequeued.

\subsection{Segment Access}

\begin{table}[t]
   \centering
   % \vspace{-4mm}
   \caption{Experiment Setup}
   \label{tab:config}
   \vspace{-4mm}
   \small
   \begin{threeparttable}
   \begin{tabular}{p{2cm} p{5.5cm}}
      \toprule[0.5pt]
       \textbf{Module} & \textbf{Configuration} \\
       \midrule
       Platform & Intel Stratix 10 GX 10M FPGA \\
       \midrule
       Scalar Core & 1 In-order, two-issue Shuttle core @ 20MHz \\
       \midrule
       Caches & Private L1 I-Cache: 16KB, 8-way\newline
               Private L1 D-Cache: 16KB, 4-way\newline
               Shared L2 cache: 512KB, 8-way, 4-bank \\
       \midrule
       Memory & 2GiB 64-bit DDR4 \\
       \midrule
       Vector Unit & P-Config: VLEN 512, DLEN 512, MLEN 512\newline
            E-Config: VLEN 256, DLEN 128, MLEN 128 \\
      \bottomrule[0.8pt]
   \end{tabular}
   \end{threeparttable}
   \vspace{-4mm}
\end{table}

Segment operations can be implemented through two distinct approaches: Segment-wise (column-wise) and Field-wise (row-wise). Here we detail these approaches in the context of Segment Loads.

The Segment-wise approach adheres to ISA semantics, where each split memory operation writes to the same segment (column). For segment loads, LAS splits operations based on segment and MLEN constraints. Memory accesses targeting the same segment within an aligned MLEN region are coalesced into a single access. After splitting, the process follows a similar request-sending pattern as strided access. When ordered responses arrive, they first enter LSDO for byte-level alignment shifting, after which LSDO writes the processed data to RCVRF using column-wise access.

The alternative Field-wise approach deviates from ISA semantics by decomposing segment operations into strided accesses or indexed accesses for each row, following the standard strided/indexed processing flow thereafter.

The performance implications of these approaches can be illustrated through an example: Consider a segment unit-stride load with base address offset=0, FIELD=2, VL=8, and EEW=8. The Segment-wise approach generates 8 memory operations, each accessing 2 bytes to write to one segment. In contrast, the Field-wise approach splits the operation into 2 strided accesses with stride=2, where each strided access generates one memory operation accessing 8 bytes. 

While EARTH's design allows for dynamic selection between these approaches based on a calculated coalescing factor to optimize performance, the current implementation exclusively uses the Segment-wise approach to maintain strict ISA semantic compliance.

\subsection{Unit-stride and Indexed Access} 
For unit-stride and indexed load, EARTH maintains Saturn's requesting process but differs in response handling: memory responses are directed to LSDO rather than LMU. LSDO performs byte-level alignment and data is written back to RCVRF in a row-wise manner.
For store operations, EARTH retrieves data from RCVRF through row-wise access, processes it through LSDO for byte-level alignment, and then initiates memory requests.

\begin{figure}[t]
    \centering
    \includegraphics[width=\linewidth]{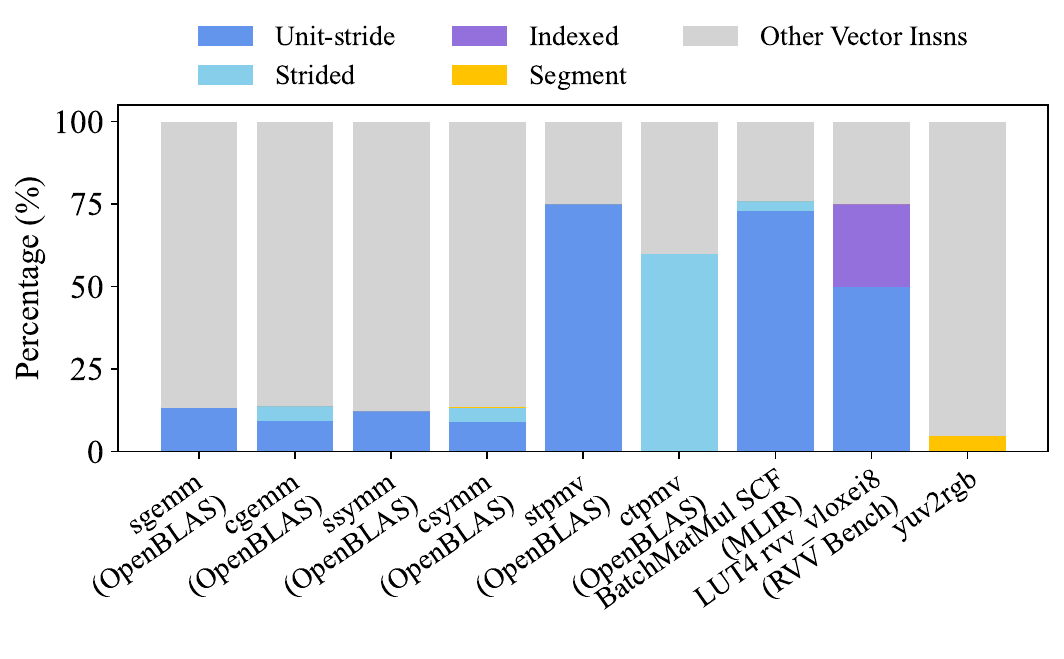}
    \vspace{-8mm}
    \caption{Vector instruction distribution}
    \label{fig:mem_dist}
    \vspace{-4mm}
\end{figure}

\begin{figure*}[t]
    \centering
    \vspace{-4mm}
    \includegraphics[width=\linewidth]{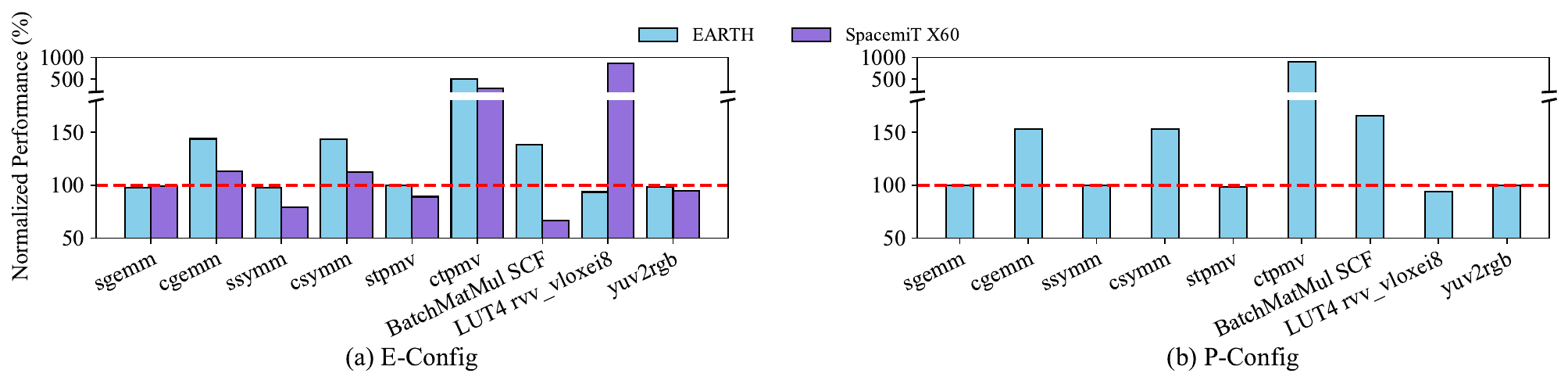}
    \vspace{-8mm}
    \caption{Diverse Pattern Benchmarks -- Normalized Performance over Saturn}
    \label{fig:perf}
    \vspace{-2mm}
\end{figure*}

\begin{figure*}[t]
    \centering
    \begin{minipage}[t]{0.48\linewidth}
        \centering
        \includegraphics[width=\linewidth]{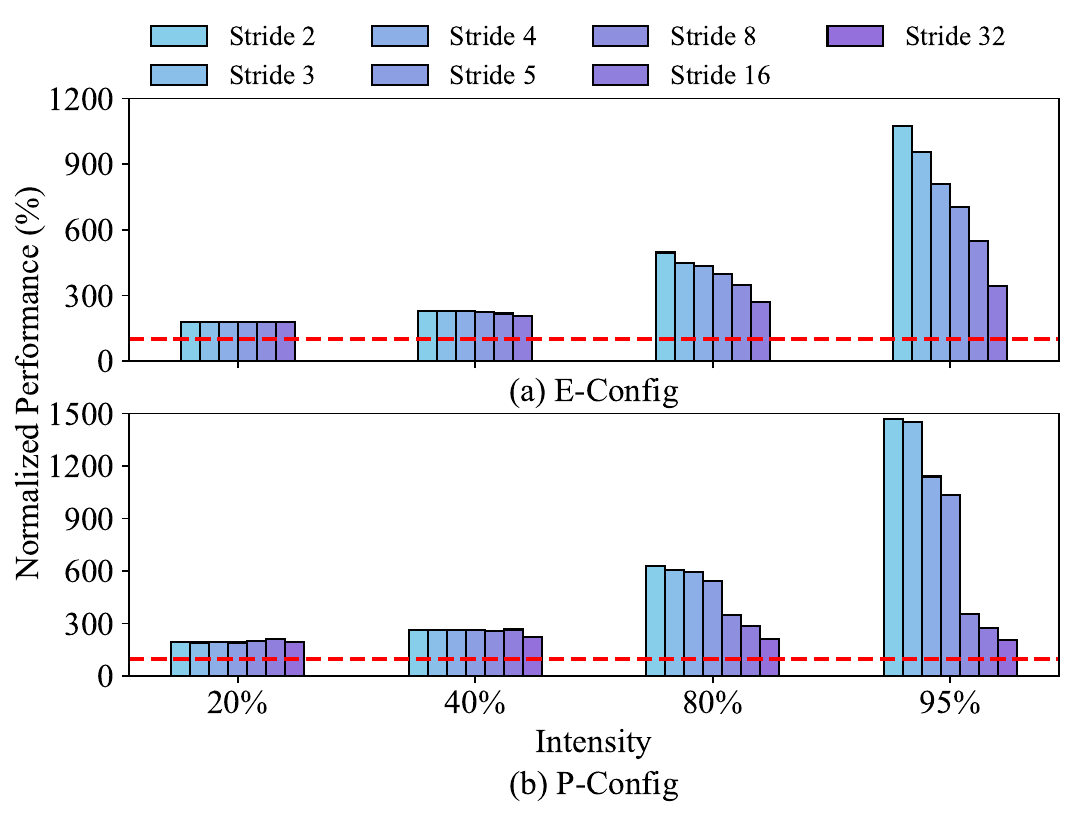}
        \vspace{-8mm}
        \caption{Strided access intensive benchmarks -- Normalized Performance Over Saturn}
        \label{fig:stride_intensive}
    \end{minipage}
    \hfill
    \begin{minipage}[t]{0.48\linewidth}
        \centering
        \includegraphics[width=\linewidth]{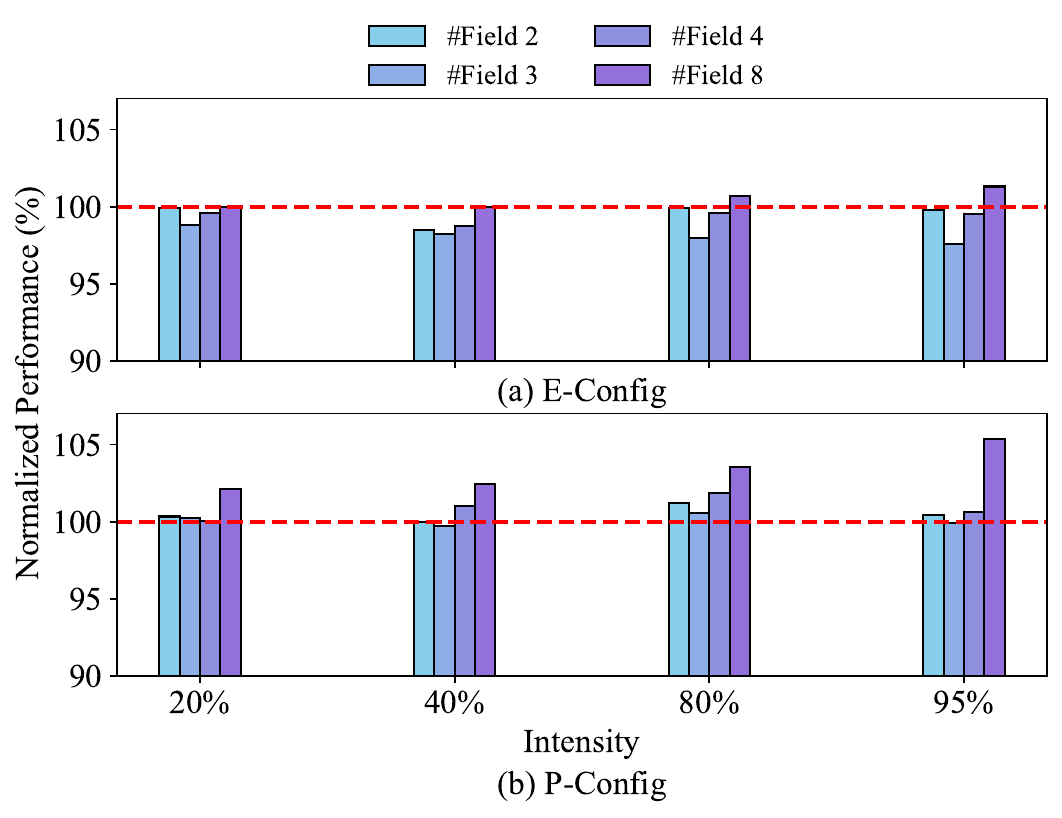}
        \vspace{-8mm}
        \caption{Segment access intensive benchmarks -- Normalized Performance Over Saturn}
        \label{fig:seg_intensive}
    \end{minipage}
    \vspace{-4mm}
\end{figure*}

\section{Evaluation}
\label{sec:evaluation}

\noindent\textbf{Settings.} We implemented EARTH in Chisel HDL and integrated it into Saturn \cite{website:saturn}. The system is integrated with a two-issue in-order Shuttle core \cite{website:shuttle}, with detailed configuration shown in Table~\ref{tab:config}. EARTH's DROM implements SSN and GSN with MLEN/8 nodes per layer across $\text{log}(\text{MLEN}/8)+1$ layers. The memory hierarchy consists of split private instruction and data L1 caches and a banked shared L2 cache serving as the last-level cache. Performance evaluation was conducted on a FPGA platform operating at 20MHz. Area measurements were obtained through Synopsys Design Compiler and power estimates were generated using Synopsys Spyglass, both with a 3-nm class process design kit and SVT cells.
 
\noindent\textbf{Workloads.} Our evaluation employs a comprehensive suite of workloads chosen to cover all vector memory access patterns. We carefully selected representative benchmarks from multiple sources: OpenBLAS \cite{website:openblas}, Buddy-MLIR Benchmark \cite{zhang2023compiler, website:buddy-benchmarks}, and RVV-Bench \cite{website:rvv-bench}. As illustrated in Figure~\ref{fig:mem_dist}, these benchmarks encompass various memory access patterns, with \textit{csymm} and \textit{yuv2rgb} demonstrating segment accesses and \textit{LUT4} exercising indexed accesses. To thoroughly evaluate EARTH's specialized features, we additionally developed stride-intensive and segment-intensive programs to evaluate performance.

\subsection{Performance: Diverse Memory Access Pattern Benchmarks}
\label{subsec:diverse}

We first evaluate performance on diverse memory access pattern benchmarks, running these benchmarks on E-Config and P-Config for both EARTH and Saturn. Additionally, we include SpacemiT Keystone K1 \cite{spacemitk1} silicon in the evaluation, which includes eight X60 cores. The X60 cores have a two-issue in-order scalar microarchitecture with a 256-VLEN vector processor, similar to our E-Config.

Figure~\ref{fig:perf} reports the performance statistics, normalized to Saturn. On benchmarks featuring only unit-stride patterns (\textit{sgemm}, \textit{ssymm}, \textit{stpmv}) \textcolor{black}{and segment patterns (\textit{yuv2rgb})}, EARTH demonstrates similar performance to Saturn on both configurations, with variations within $\pm 3\%$. On \textit{LUT4}, EARTH experiences slight performance degradation (-6.5\% and -6.1\% on E-Config and P-Config, respectively) over Saturn, due to increased pipeline stages for indexed instructions. However, EARTH demonstrates significant performance improvements on benchmarks featuring strided access patterns: \textit{cgemm} (+43.8\%, +53.3\%), \textit{csymm} (+43.6\%, +52.9\%), \textit{ctpmv} (+401.1\%, +797.2\%), and \textit{BatchMatMul SCF} (+38.5\%, +65.7\%) on E-Config and P-Config.

For comparisons with SpacemiT X60, we scale its performance by frequency ratio over EARTH. \textcolor{black}{To ensure a fair comparison, we use EARTH's E-Config which matches X60's VLEN, and reduce X60's frequency to 614.4MHz to minimize memory latency effects. While differences in architectural details and memory subsystem configurations may introduce comparison bias, we believe this methodology provides meaningful insights.} EARTH demonstrates superior performance across most benchmarks, though SpacemiT X60 achieves exceptional performance gains (+761.1\%) on \textit{LUT4}, which heavily utilizes indexed load/store operations. 
This indicates potential for future optimization of EARTH's indexed operations.

\vspace{-2mm}

\subsection{Performance: Pattern Intensive Benchmarks}
\label{subsec:intensive}

We construct stride-intensive and segment-intensive benchmark programs to evaluate the performance of EARTH and Saturn. The intensity of these benchmarks is defined as the ratio of strided or segmented instructions to the total number of vector instructions. Experiments were conducted on both E-Config and P-Config configurations under four intensity levels: \textcolor{black}{20\%, 40\%, 80\%, and 95\%}, with stride values ranging from 2 to \(\text{MLEN}/2\) for strided access and field values ranging from 2 to 8 for segment access.

\textcolor{black}{
Figure~\ref{fig:stride_intensive} presents the normalized performance of EARTH compared to Saturn on stride-intensive benchmarks. Across all configurations, EARTH demonstrates substantial performance improvements, reaching up to 14x speedup over Saturn. For P-Config, EARTH achieves an average performance improvement of 4.4x across all intensity levels and stride values, while for E-Config, the average improvement is 3.8x. EARTH's performance gains become more pronounced as benchmark intensity increases. For instance, in P-Config with a stride value of 2, EARTH achieves a 1.9x speedup at 20\% intensity, which grows significantly to 14.7x at 95\% intensity. EARTH also exhibits robust performance across varying stride values, with a clear pattern emerging: benchmarks with smaller strides consistently show higher performance improvements due to increased opportunities for memory request coalescing. For example, in E-Config at 95\% intensity, EARTH achieves a 3.4x speedup for stride=16, whereas this increases to 10.8x for stride=2. Furthermore, P-Config generally outperforms E-Config across all test cases, primarily due to its larger MLEN, which enables more effective memory coalescing operations.
}

\textcolor{black}{
Figure~\ref{fig:seg_intensive} compares EARTH's performance against Saturn on segment-intensive benchmarks. EARTH maintains comparable performance across all configurations, achieving 1.01x and 0.99x of Saturn's performance for P-Config and E-Config respectively. These results demonstrate that EARTH's elimination of segment buffers successfully achieves efficient segment handling without performance degradation, while reducing hardware costs.
}

\subsection{Area Analysis}

\begin{figure}[t]
    \centering
    % \vspace{-2mm}
    \includegraphics[width=\linewidth]{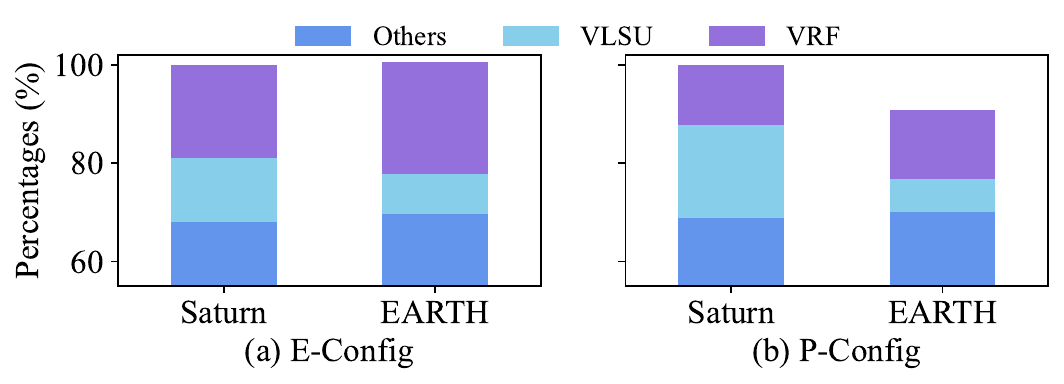}
    \vspace{-6mm}
    \caption{Area Distribution -- Normalized to Saturn's Area}
    \label{fig:area}
    \vspace{-4mm}
\end{figure}

We estimate the area overhead using Synopsys Design Compiler. Figure~\ref{fig:area} presents the area distribution of EARTH and Saturn, normalized to Saturn's total area.

EARTH's RCVRF increases the VRF area due to the incorporation of the DROM and Block Shifters. In E-Config, the VRF area increases by 20.35\%, while in P-Config, the increase is reduced to 15.15\%. In contrast, EARTH significantly reduces the VLSU area by eliminating segment buffers. For E-Config, this results in a 37.25\% reduction in VLSU area, while for P-Config, the reduction is a substantial 64.71\%.

In E-Config, due to the need to integrate EARTH with Saturn’s original structure, additional area is required in other modules. As a result, despite reductions in VLSU and VRF areas, E-Config exhibits a slight overall area increase of 0.58\%. In contrast, for P-Config, which suffers from segment buffer explosion in Saturn, EARTH achieves a significant total area reduction of 9.11\%.

\subsection{Power Analysis}

We conduct a comprehensive power analysis using Synopsys SpyGlass to evaluate EARTH's energy efficiency, focusing on the strided and segment access patterns, as these are the primary patterns optimized by EARTH. 

For each memory access pattern, we utilized all program snippets of the relevant instructions from riscv-vector-tests \cite{website:riscv-vector-tests}. We used waveforms for load and store operations with different ELEN values ranging from 8 to 64 as activity data references. We then calculated the average power consumption of each pattern as the result.

\begin{figure}[t]
    \centering
    % \vspace{-3mm}
    \includegraphics[width=\linewidth]{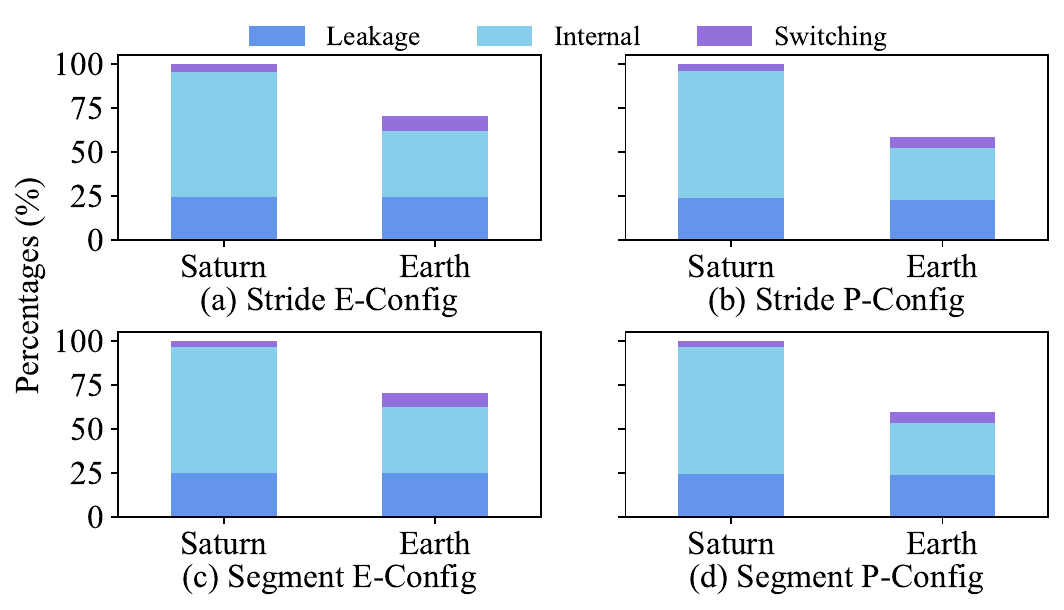}
    \vspace{-6mm}
    \caption{Power Consumption Distribution -- Normalized to Saturn's power}
    \label{fig:power}
    \vspace{-6mm}
\end{figure}

Figure~\ref{fig:power} presents the power consumption distribution of EARTH and Saturn, normalized to Saturn's total power. The power consumption is divided into three components: leakage, internal, and switching power. While EARTH achieves significant reductions in internal power and maintains comparable leakage power with Saturn. This increased switching power originates from EARTH's more aggressive shifting logic, which enables better performance but requires more signal transitions.

Despite the switching power overhead, EARTH achieves a net power reduction of 29.4--29.7\% compared to Saturn on E-Config and 40.3--41.6\% on P-Config. These savings are primarily due to substantial reductions in internal power consumption, driven by two key architectural innovations: (1) the stride-aware coalescing mechanism, which reduces the total number of strided memory requests, eliminating redundant memory traffic and associated control logic activities, and (2) the removal of the dedicated segment buffers required in Saturn, significantly reducing buffer maintenance overhead. The consistent power reduction across both access patterns (29.7\% and 41.6\% for strided accesses, 29.4\% and 40.3\% for segment accesses) demonstrates EARTH's robust energy efficiency across diverse memory access behaviors.

\section{Conclusion}
\label{sec:conclusion}

In this paper, we detailed the design and implementation of EARTH, an efficient architecture for RISC-V vector memory access patterns. We introduced DROM, LSDO, and RCVRF, optimizations that enable coalesced strided instruction memory access and buffer-free segment instruction processing. By implementing these optimizations on Saturn, specifically a modern in-order two-issue RISC-V CPU with a Vector Unit fully compliant with the RISC-V Vector 1.0 specification, we provide a foundation for further exploration and research. This implementation allows for the use and optimization of vector load/store instructions in both hardware and applications. Our evaluation demonstrates that our approach offers comparable, and in some cases superior, performance and area advantages over existing open-source and commercial solutions. We believe that the overall architecture can serve as a design paradigm, providing efficient memory access support for computing data flow innovation on the RISC-V architecture. 
\textcolor{black}{EARTH's architecture inherently supports scalability. While our current prototype employs a single LSU, the design naturally enables GPU-style multi-LSU configurations. This scalability pathway allows future implementations to exploit memory-level parallelism more aggressively, mirroring the trajectory of modern GPU architectures.  }

\bibliographystyle{ACM-Reference-Format}
\bibliography{acmart}


\begin{thebibliography}{28}

%%% the Web address.

\ifx \showCODEN    \undefined \def \showCODEN     #1{\unskip}     \fi
\ifx \showDOI      \undefined \def \showDOI       #1{#1}\fi
\ifx \showISBNx    \undefined \def \showISBNx     #1{\unskip}     \fi
\ifx \showISBNxiii \undefined \def \showISBNxiii  #1{\unskip}     \fi
\ifx \showISSN     \undefined \def \showISSN      #1{\unskip}     \fi
\ifx \showLCCN     \undefined \def \showLCCN      #1{\unskip}     \fi
\ifx \shownote     \undefined \def \shownote      #1{#1}          \fi
\ifx \showarticletitle \undefined \def \showarticletitle #1{#1}   \fi
\ifx \showURL      \undefined \def \showURL       {\relax}        \fi
% invisible to TeX
\providecommand\bibfield[2]{#2}
\providecommand\bibinfo[2]{#2}
\providecommand\natexlab[1]{#1}
\providecommand\showeprint[2][]{arXiv:#2}

\bibitem[Alliance(2024)]%
        {website:t1}
\bibfield{author}{\bibinfo{person}{CHIPS Alliance}.}
  \bibinfo{year}{2024}\natexlab{}.
\newblock \bibinfo{title}{T1: A RISC-V Core}.
\newblock
\newblock
\urldef\tempurl%
\url{https://github.com/chipsalliance/t1}
\showURL{%
\tempurl}


\bibitem[ARM(2023)]%
        {arm_ddi0487}
\bibfield{author}{\bibinfo{person}{ARM}.} \bibinfo{year}{2023}\natexlab{}.
\newblock \bibinfo{title}{ARM Architecture Reference Manual for ARMv8-A}.
\newblock
\newblock
\urldef\tempurl%
\url{https://developer.arm.com/documentation/ddi0487/latest}
\showURL{%
\tempurl}


\bibitem[Bachrach et~al\mbox{.}(2012)]%
        {jonathan2012chisel}
\bibfield{author}{\bibinfo{person}{Jonathan Bachrach}, \bibinfo{person}{Huy
  Vo}, \bibinfo{person}{Brian Richards}, \bibinfo{person}{Yunsup Lee},
  \bibinfo{person}{Andrew Waterman}, \bibinfo{person}{Rimas Avi\v{z}ienis},
  \bibinfo{person}{John Wawrzynek}, {and} \bibinfo{person}{Krste
  Asanovi\'{c}}.} \bibinfo{year}{2012}\natexlab{}.
\newblock \showarticletitle{Chisel: constructing hardware in a Scala embedded
  language}. In \bibinfo{booktitle}{\emph{Proceedings of the 49th Annual Design
  Automation Conference}} (San Francisco, California)
  \emph{(\bibinfo{series}{DAC '12})}. \bibinfo{publisher}{Association for
  Computing Machinery}, \bibinfo{address}{New York, NY, USA},
  \bibinfo{pages}{1216–1225}.
\newblock
\showISBNx{9781450311991}
\urldef\tempurl%
\url{https://doi.org/10.1145/2228360.2228584}
\showDOI{\tempurl}


\bibitem[Boemer et~al\mbox{.}(2021)]%
        {boemer2021intel}
\bibfield{author}{\bibinfo{person}{Fabian Boemer}, \bibinfo{person}{Sejun Kim},
  \bibinfo{person}{Gelila Seifu}, \bibinfo{person}{Fillipe DM~de Souza}, {and}
  \bibinfo{person}{Vinodh Gopal}.} \bibinfo{year}{2021}\natexlab{}.
\newblock \showarticletitle{Intel HEXL: accelerating homomorphic encryption
  with Intel AVX512-IFMA52}. In \bibinfo{booktitle}{\emph{Proceedings of the
  9th on Workshop on Encrypted Computing \& Applied Homomorphic Cryptography}}.
  \bibinfo{pages}{57--62}.
\newblock


\bibitem[Cavalcante et~al\mbox{.}(2019)]%
        {cavalcante2019ara}
\bibfield{author}{\bibinfo{person}{Matheus Cavalcante}, \bibinfo{person}{Fabian
  Schuiki}, \bibinfo{person}{Florian Zaruba}, \bibinfo{person}{Michael
  Schaffner}, {and} \bibinfo{person}{Luca Benini}.}
  \bibinfo{year}{2019}\natexlab{}.
\newblock \showarticletitle{Ara: A 1-GHz+ scalable and energy-efficient RISC-V
  vector processor with multiprecision floating-point support in 22-nm FD-SOI}.
\newblock \bibinfo{journal}{\emph{IEEE Transactions on Very Large Scale
  Integration (VLSI) Systems}} \bibinfo{volume}{28}, \bibinfo{number}{2}
  (\bibinfo{year}{2019}), \bibinfo{pages}{530--543}.
\newblock


\bibitem[Coder(2024)]%
        {website:rvv-bench}
\bibfield{author}{\bibinfo{person}{Camel Coder}.}
  \bibinfo{year}{2024}\natexlab{}.
\newblock \bibinfo{booktitle}{\emph{RISC-V Vector benchmark}}.
\newblock
\urldef\tempurl%
\url{https://github.com/camel-cdr/rvv-bench}
\showURL{%
\tempurl}


\bibitem[Contributors(2024a)]%
        {website:buddy-benchmarks}
\bibfield{author}{\bibinfo{person}{Buddy-Compiler Contributors}.}
  \bibinfo{year}{2024}\natexlab{a}.
\newblock \bibinfo{booktitle}{\emph{Buddy Benchmark}}.
\newblock
\urldef\tempurl%
\url{https://github.com/buddy-compiler/buddy-benchmark}
\showURL{%
\tempurl}


\bibitem[Contributors(2024b)]%
        {website:riscv-vector-tests}
\bibfield{author}{\bibinfo{person}{CHIPS~Alliance Contributors}.}
  \bibinfo{year}{2024}\natexlab{b}.
\newblock \bibinfo{booktitle}{\emph{RISC-V Vector Tests Generator}}.
\newblock
\urldef\tempurl%
\url{https://github.com/chipsalliance/riscv-vector-tests}
\showURL{%
\tempurl}


\bibitem[Contributors(2024c)]%
        {website:openblas}
\bibfield{author}{\bibinfo{person}{OpenBLAS Contributors}.}
  \bibinfo{year}{2024}\natexlab{c}.
\newblock \bibinfo{booktitle}{\emph{OpenBLAS: An optimized BLAS library}}.
\newblock
\urldef\tempurl%
\url{https://github.com/OpenMathLib/OpenBLAS}
\showURL{%
\tempurl}


\bibitem[Contributors(2024d)]%
        {website:shuttle}
\bibfield{author}{\bibinfo{person}{UCB-BAR Contributors}.}
  \bibinfo{year}{2024}\natexlab{d}.
\newblock \bibinfo{booktitle}{\emph{Shuttle: A Rocket-based Superscalar
  In-order RISC-V Core}}.
\newblock
\urldef\tempurl%
\url{https://github.com/ucb-bar/shuttle}
\showURL{%
\tempurl}


\bibitem[Corporation(2023a)]%
        {intel_sdm}
\bibfield{author}{\bibinfo{person}{Intel Corporation}.}
  \bibinfo{year}{2023}\natexlab{a}.
\newblock \bibinfo{title}{Intel® 64 and IA-32 Architectures Software
  Developer’s Manual: Combined Volumes 2A, 2B, 2C, and 2D: Instruction Set
  Reference, A-Z}.
\newblock
\newblock
\urldef\tempurl%
\url{https://www.intel.com/content/www/us/en/content-details/835757/intel-64-and-ia-32-architectures-software-developer-s-manual-combined-volumes-2a-2b-2c-and-2d-instruction-set-reference-a-z.html}
\showURL{%
\tempurl}


\bibitem[Corporation(2023b)]%
        {intel_avx512}
\bibfield{author}{\bibinfo{person}{Intel Corporation}.}
  \bibinfo{year}{2023}\natexlab{b}.
\newblock \bibinfo{title}{Intel® Advanced Vector Extensions 512 (Intel®
  AVX-512) Overview}.
\newblock
\newblock
\urldef\tempurl%
\url{https://www.intel.com/content/www/us/en/architecture-and-technology/avx-512-overview.html}
\showURL{%
\tempurl}


\bibitem[Crago et~al\mbox{.}(2018)]%
        {gpu}
\bibfield{author}{\bibinfo{person}{Neal~C. Crago}, \bibinfo{person}{Mark
  Stephenson}, {and} \bibinfo{person}{Stephen~W. Keckler}.}
  \bibinfo{year}{2018}\natexlab{}.
\newblock \showarticletitle{Exposing Memory Access Patterns to Improve
  Instruction and Memory Efficiency in GPUs}.
\newblock \bibinfo{journal}{\emph{ACM Trans. Archit. Code Optim.}}
  \bibinfo{volume}{15}, \bibinfo{number}{4}, Article \bibinfo{articleno}{45}
  (\bibinfo{date}{Oct.} \bibinfo{year}{2018}), \bibinfo{numpages}{23}~pages.
\newblock
\showISSN{1544-3566}
\urldef\tempurl%
\url{https://doi.org/10.1145/3280851}
\showDOI{\tempurl}


\bibitem[Flynn(1972)]%
        {flynn1972some}
\bibfield{author}{\bibinfo{person}{Michael~J Flynn}.}
  \bibinfo{year}{1972}\natexlab{}.
\newblock \showarticletitle{Some computer organizations and their
  effectiveness}.
\newblock \bibinfo{journal}{\emph{IEEE transactions on computers}}
  \bibinfo{volume}{100}, \bibinfo{number}{9} (\bibinfo{year}{1972}),
  \bibinfo{pages}{948--960}.
\newblock


\bibitem[Gathu(2024)]%
        {gathu2024high}
\bibfield{author}{\bibinfo{person}{Simon Gathu}.}
  \bibinfo{year}{2024}\natexlab{}.
\newblock \showarticletitle{High-Performance Computing and Big Data: Emerging
  Trends in Advanced Computing Systems for Data-Intensive Applications}.
\newblock \bibinfo{journal}{\emph{Journal of Advanced Computing Systems}}
  \bibinfo{volume}{4}, \bibinfo{number}{8} (\bibinfo{year}{2024}),
  \bibinfo{pages}{22--35}.
\newblock


\bibitem[Guo et~al\mbox{.}(2023)]%
        {guo2023olive}
\bibfield{author}{\bibinfo{person}{Cong Guo}, \bibinfo{person}{Jiaming Tang},
  \bibinfo{person}{Weiming Hu}, \bibinfo{person}{Jingwen Leng},
  \bibinfo{person}{Chen Zhang}, \bibinfo{person}{Fan Yang},
  \bibinfo{person}{Yunxin Liu}, \bibinfo{person}{Minyi Guo}, {and}
  \bibinfo{person}{Yuhao Zhu}.} \bibinfo{year}{2023}\natexlab{}.
\newblock \showarticletitle{Olive: Accelerating large language models via
  hardware-friendly outlier-victim pair quantization}. In
  \bibinfo{booktitle}{\emph{ISCA}}. \bibinfo{pages}{1--15}.
\newblock


\bibitem[Hennessy and Patterson(2017)]%
        {hennessy2017computer}
\bibfield{author}{\bibinfo{person}{John~L Hennessy} {and}
  \bibinfo{person}{David~A Patterson}.} \bibinfo{year}{2017}\natexlab{}.
\newblock \bibinfo{booktitle}{\emph{Computer architecture: a quantitative
  approach}}.
\newblock


\bibitem[Hua et~al\mbox{.}(2023)]%
        {hua2023edge}
\bibfield{author}{\bibinfo{person}{Haochen Hua}, \bibinfo{person}{Yutong Li},
  \bibinfo{person}{Tonghe Wang}, \bibinfo{person}{Nanqing Dong},
  \bibinfo{person}{Wei Li}, {and} \bibinfo{person}{Junwei Cao}.}
  \bibinfo{year}{2023}\natexlab{}.
\newblock \showarticletitle{Edge computing with artificial intelligence: A
  machine learning perspective}.
\newblock \bibinfo{journal}{\emph{Comput. Surveys}} \bibinfo{volume}{55},
  \bibinfo{number}{9} (\bibinfo{year}{2023}), \bibinfo{pages}{1--35}.
\newblock


\bibitem[Perotti et~al\mbox{.}(2022)]%
        {perotti2022new}
\bibfield{author}{\bibinfo{person}{Matteo Perotti}, \bibinfo{person}{Matheus
  Cavalcante}, \bibinfo{person}{Nils Wistoff}, \bibinfo{person}{Renzo Andri},
  \bibinfo{person}{Lukas Cavigelli}, {and} \bibinfo{person}{Luca Benini}.}
  \bibinfo{year}{2022}\natexlab{}.
\newblock \showarticletitle{A “new ara” for vector computing: An open
  source highly efficient risc-v v 1.0 vector processor design}. In
  \bibinfo{booktitle}{\emph{ASAP}}. IEEE, \bibinfo{pages}{43--51}.
\newblock


\bibitem[{RISC-V International}(2021)]%
        {rvvspec}
\bibfield{author}{\bibinfo{person}{{RISC-V International}}.}
  \bibinfo{year}{2021}\natexlab{}.
\newblock \bibinfo{title}{{RISC-V} {V}ector {E}xtension {V}ersion 1.0}.
\newblock
\newblock
\urldef\tempurl%
\url{https://github.com/riscv/riscv-v-spec}
\showURL{%
\tempurl}


\bibitem[Russell(1978)]%
        {russell1978cray}
\bibfield{author}{\bibinfo{person}{Richard~M Russell}.}
  \bibinfo{year}{1978}\natexlab{}.
\newblock \showarticletitle{The CRAY-1 computer system}.
\newblock \bibinfo{journal}{\emph{Commun. ACM}} \bibinfo{volume}{21},
  \bibinfo{number}{1} (\bibinfo{year}{1978}), \bibinfo{pages}{63--72}.
\newblock


\bibitem[{SpacemiT Technology}(2024)]%
        {spacemitk1}
\bibfield{author}{\bibinfo{person}{{SpacemiT Technology}}.}
  \bibinfo{year}{2024}\natexlab{}.
\newblock \bibinfo{booktitle}{\emph{SpacemiT Key Stone K1}}.
\newblock SpacemiT Technology.
\newblock
\urldef\tempurl%
\url{https://www.spacemit.com/en/key-stone-k1/}
\showURL{%
\tempurl}


\bibitem[Stephens et~al\mbox{.}(2017)]%
        {stephens2017arm}
\bibfield{author}{\bibinfo{person}{Nigel Stephens}, \bibinfo{person}{Stuart
  Biles}, \bibinfo{person}{Matthias Boettcher}, \bibinfo{person}{Jacob Eapen},
  \bibinfo{person}{Mbou Eyole}, \bibinfo{person}{Giacomo Gabrielli},
  \bibinfo{person}{Matt Horsnell}, \bibinfo{person}{Grigorios Magklis},
  \bibinfo{person}{Alejandro Martinez}, \bibinfo{person}{Nathanael Premillieu},
  {et~al\mbox{.}}} \bibinfo{year}{2017}\natexlab{}.
\newblock \showarticletitle{The ARM scalable vector extension}.
\newblock \bibinfo{journal}{\emph{IEEE micro}} \bibinfo{volume}{37},
  \bibinfo{number}{2} (\bibinfo{year}{2017}), \bibinfo{pages}{26--39}.
\newblock


\bibitem[Wang et~al\mbox{.}(2024)]%
        {wang2024xiangshan}
\bibfield{author}{\bibinfo{person}{Kaifan Wang}, \bibinfo{person}{Jian Chen},
  \bibinfo{person}{Yinan Xu}, \bibinfo{person}{Zihao Yu},
  \bibinfo{person}{Zifei Zhang}, \bibinfo{person}{Guokai Chen},
  \bibinfo{person}{Xuan Hu}, \bibinfo{person}{Linjuan Zhang},
  \bibinfo{person}{Xi Chen}, \bibinfo{person}{Wei He}, {et~al\mbox{.}}}
  \bibinfo{year}{2024}\natexlab{}.
\newblock \showarticletitle{XiangShan: An Open-Source Project for
  High-Performance RISC-V Processors Meeting Industrial-Grade Standards}. In
  \bibinfo{booktitle}{\emph{HCS}}. IEEE Computer Society,
  \bibinfo{pages}{1--25}.
\newblock


\bibitem[Zhang et~al\mbox{.}(2023a)]%
        {zhang2023axi}
\bibfield{author}{\bibinfo{person}{Chi Zhang}, \bibinfo{person}{Paul
  Scheffler}, \bibinfo{person}{Thomas Benz}, \bibinfo{person}{Matteo Perotti},
  {and} \bibinfo{person}{Luca Benini}.} \bibinfo{year}{2023}\natexlab{a}.
\newblock \showarticletitle{AXI-pack: Near-memory bus packing for
  bandwidth-efficient irregular workloads}. In
  \bibinfo{booktitle}{\emph{DATE}}. IEEE, \bibinfo{pages}{1--6}.
\newblock


\bibitem[Zhang et~al\mbox{.}(2024)]%
        {zhang2024near}
\bibfield{author}{\bibinfo{person}{Chi Zhang}, \bibinfo{person}{Paul
  Scheffler}, \bibinfo{person}{Thomas Benz}, \bibinfo{person}{Matteo Perotti},
  {and} \bibinfo{person}{Luca Benini}.} \bibinfo{year}{2024}\natexlab{}.
\newblock \showarticletitle{Near-Memory Parallel Indexing and Coalescing:
  Enabling Highly Efficient Indirect Access for SpMV}. In
  \bibinfo{booktitle}{\emph{DATE}}. IEEE, \bibinfo{pages}{1--6}.
\newblock


\bibitem[Zhang et~al\mbox{.}(2023b)]%
        {zhang2023compiler}
\bibfield{author}{\bibinfo{person}{Hongbin Zhang}, \bibinfo{person}{Mingjie
  Xing}, \bibinfo{person}{Yanjun Wu}, {and} \bibinfo{person}{Chen Zhao}.}
  \bibinfo{year}{2023}\natexlab{b}.
\newblock \showarticletitle{Compiler Technologies in Deep Learning Co-Design: A
  Survey}.
\newblock \bibinfo{journal}{\emph{Intelligent Computing}}
  (\bibinfo{year}{2023}).
\newblock


\bibitem[Zhao et~al\mbox{.}(2024)]%
        {website:saturn}
\bibfield{author}{\bibinfo{person}{Jerry Zhao}, \bibinfo{person}{Daniel Grubb},
  \bibinfo{person}{Miles Rusch}, \bibinfo{person}{Tianrui Wei},
  \bibinfo{person}{Kevin Anderson}, \bibinfo{person}{Borivoje Nikolic}, {and}
  \bibinfo{person}{Krste Asanović}.} \bibinfo{year}{2024}\natexlab{}.
\newblock \bibinfo{booktitle}{\emph{The Saturn Microarchitecture Manual}}.
\newblock \bibinfo{type}{{T}echnical {R}eport} UCB/EECS-2024-215.
  \bibinfo{institution}{EECS Department, University of California, Berkeley}.
\newblock
\urldef\tempurl%
\url{http://www2.eecs.berkeley.edu/Pubs/TechRpts/2024/EECS-2024-215.html}
\showURL{%
\tempurl}


\end{thebibliography}

\end{document}